\documentclass[iop]{emulateapj}
\pdfoutput=1
\usepackage{mathtools}
\usepackage{hyperref}
\usepackage{amsmath, amssymb}
\usepackage{url}
\usepackage{natbib}
\usepackage{color}
\usepackage{float}

\newcommand{\lr}[1]{\left(#1\right)}
\newcommand{\LR}[1]{\left[#1\right]}
\newcommand{\Lr}[1]{\left\{#1\right\}}
\newcommand{\E}{\times 10^}
\newcommand{\unt}[1]{\mathrm{\;#1}}

\newcommand{\ssub}[1]{_{\text{\scriptsize{#1}}}}

\newcommand{\snana}{{\tt SNANA}}
\newcommand{\galapagos}{{G\textsc{alapagos}}}

\newcommand{\ddlr}{$d_\text{DLR}$}
\newcommand{\sersic}{S{\'e}rsic}
\newcommand{\hst}{\textit{HST}}

\slugcomment{Accepted by AJ}

\shortauthors{Gupta et al.}

\begin{document}

\title{Host Galaxy Identification for Supernova Surveys}

\author{
	Ravi R. Gupta\altaffilmark{1}, 
	Steve Kuhlmann\altaffilmark{1}, 
	Eve Kovacs\altaffilmark{1}, 
	Harold Spinka\altaffilmark{1}, 
	Richard Kessler\altaffilmark{6,3},
	Daniel A. Goldstein\altaffilmark{4,5},
	Camille Liotine\altaffilmark{1}, 
	Katarzyna Pomian\altaffilmark{1},  
	Chris B. D'Andrea\altaffilmark{6,7},  
	Mark Sullivan\altaffilmark{7}, 
	Jorge Carretero\altaffilmark{8,9}, 
	Francisco J. Castander\altaffilmark{8}, 
	Robert C. Nichol\altaffilmark{6}, 
	David A. Finley\altaffilmark{10}, 
	John A. Fischer\altaffilmark{11}, 
	Ryan J. Foley\altaffilmark{12,13}, 
	Alex G. Kim\altaffilmark{5}, 
	Andreas Papadopoulos\altaffilmark{6,14}, 
	Masao Sako\altaffilmark{11}, 
	Daniel M. Scolnic\altaffilmark{6}, 
	Mathew Smith\altaffilmark{7}, 
	Brad E. Tucker\altaffilmark{15}, 
	Syed Uddin\altaffilmark{17}, 
	Rachel C. Wolf\altaffilmark{11}, 
	Fang Yuan\altaffilmark{15,16},
	Tim M. C. Abbott\altaffilmark{18}, 
	Filipe B. Abdalla\altaffilmark{19,20}, 
	Aur{\'e}lien Benoit-L{\'e}vy\altaffilmark{21,19,22}, 
	Emmanuel Bertin\altaffilmark{21,22}, 
	David Brooks\altaffilmark{19}, 
	Aurelio Carnero Rosell\altaffilmark{23,24}, 
	Matias Carrasco Kind\altaffilmark{12,25}, 
	Carlos E. Cunha\altaffilmark{26}, 
	Luiz N. da Costa\altaffilmark{23,24}, 
	Shantanu Desai\altaffilmark{27,28}, 
	Peter Doel\altaffilmark{19}, 
	Tim F. Eifler\altaffilmark{11,29}, 
	August E. Evrard\altaffilmark{30,31}, 
	Brenna Flaugher\altaffilmark{10}, 
	Pablo Fosalba\altaffilmark{8}, 
	Enrique Gazta\~naga\altaffilmark{8}, 
	Daniel Gruen\altaffilmark{26,32}, 
	Robert Gruendl\altaffilmark{12,25}, 
	David J. James\altaffilmark{18}, 
	Kyler Kuehn\altaffilmark{33}, 
	Nikolay Kuropatkin\altaffilmark{10}, 
	Marcio A. G. Maia\altaffilmark{23,24}, 
	Jennifer L. Marshall\altaffilmark{34}, 
	Ramon Miquel\altaffilmark{35,9}, 
	Andr{\'e}s A. Plazas\altaffilmark{29}, 
	A. Kathy Romer\altaffilmark{36}, 
	Eusebio S{\'a}nchez\altaffilmark{37}, 
	Michael Schubnell\altaffilmark{31},
	Ignacio Sevilla-Noarbe\altaffilmark{37},
	Fl{\'a}via Sobreira\altaffilmark{23},
	Eric Suchyta\altaffilmark{11}, 
	Molly E. C. Swanson\altaffilmark{25},
	Gregory Tarle\altaffilmark{31}, 
	Alistair R. Walker\altaffilmark{18},
	William Wester\altaffilmark{10}
}

\email{raviryan@gmail.com}

\altaffiltext{1}{
  Argonne National Laboratory, 
  9700 South Cass Avenue, Lemont, IL 60439, USA
}
\altaffiltext{2}{
  Kavli Institute for Cosmological Physics, 
  University of Chicago, Chicago, IL 60637, USA
}
\altaffiltext{3}{
  Department of Astronomy and Astrophysics, 
  University of Chicago, 5640 South Ellis Avenue, 
  Chicago, IL 60637, USA
}
\altaffiltext{4}{
  Department of Astronomy, University of California, 
  Berkeley, 501 Campbell Hall \#3411, Berkeley, CA 94720
}
\altaffiltext{5}{
  Physics Division, Lawrence Berkeley National Laboratory, 
  1 Cyclotron Road, Berkeley, CA 94720, USA
}
\altaffiltext{6}{
  Institute of Cosmology and Gravitation, 
  University of Portsmouth, Portsmouth, PO1 3FX, UK
}
\altaffiltext{7}{
  Department of Physics and Astronomy, 
  University of Southampton, Southampton, SO17 1BJ, UK
}
\altaffiltext{8}{
  Institut de Ci\`encies de l'Espai, IEEC-CSIC, 
  Campus UAB, Carrer de Can Magrans, s/n,  08193 Bellaterra, Barcelona, Spain
}
\altaffiltext{9}{
  Institut de F\'isica d'Altes Energies (IFAE), 
  The Barcelona Institute of Science and Technology, 
  Campus UAB, 08193 Bellaterra (Barcelona) Spain
}
\altaffiltext{10}{
  Fermi National Accelerator Laboratory, 
  P. O. Box 500, Batavia, IL 60510, USA
}
\altaffiltext{11}{
  Department of Physics and Astronomy,
  University of Pennsylvania, 209 South 33rd Street,
  Philadelphia, PA 19104, USA
}
\altaffiltext{12}{
  Department of Astronomy, 
  University of Illinois, 1002 W. Green Street, 
  Urbana, IL 61801, USA
}
\altaffiltext{13}{
  Department of Physics, 
  University of Illinois, 1110 W. Green Street, 
  Urbana, IL 61801, USA
}
\altaffiltext{14}{
  School of Sciences, European University Cyprus, 
  6 Diogenis Str., Engomi, 1516 Nicosia, Cyprus
}
\altaffiltext{15}{
  The Research School of Astronomy and Astrophysics, 
  Australian National University, Mount Stromlo Observatory, 
  via Cotter Road, Weston Creek, ACT 2611, Australia
}
\altaffiltext{16}{
  ARC Centre of Excellence for All-sky Astrophysics (CAASTRO)
}
\altaffiltext{17}{
  Centre for Astrophysics \& Supercomputing, 
  Swinburne University of Technology, Victoria 3122, Australia
}
\altaffiltext{18}{
  Cerro Tololo Inter-American Observatory, 
  National Optical Astronomy Observatory, 
  Casilla 603, La Serena, Chile
}
\altaffiltext{19}{
  Department of Physics \& Astronomy, University College London, 
  Gower Street, London, WC1E 6BT, UK
}
\altaffiltext{20}{
  Department of Physics and Electronics, Rhodes University, 
  PO Box 94, Grahamstown, 6140, South Africa
}
\altaffiltext{21}{
  CNRS, UMR 7095, Institut d'Astrophysique de Paris, F-75014, Paris, France
}
\altaffiltext{22}{
  Sorbonne Universit\'es, UPMC Univ Paris 06, UMR 7095, 
  Institut d'Astrophysique de Paris, F-75014, Paris, France
}
\altaffiltext{23}{
  Laborat\'orio Interinstitucional de e-Astronomia - LIneA, 
  Rua Gal. Jos\'e Cristino 77, Rio de Janeiro, RJ - 20921-400, Brazil
}
\altaffiltext{24}{
  Observat\'orio Nacional, Rua Gal. Jos\'e Cristino 77, 
  Rio de Janeiro, RJ - 20921-400, Brazil
}
\altaffiltext{25}{
  National Center for Supercomputing Applications, 
  1205 West Clark St., Urbana, IL 61801, USA
}
\altaffiltext{26}{
  Kavli Institute for Particle Astrophysics \& Cosmology, 
  P. O. Box 2450, Stanford University, Stanford, CA 94305, USA
}
\altaffiltext{27}{
  Excellence Cluster Universe, Boltzmannstr.\ 2, 85748 Garching, Germany
}
\altaffiltext{28}{
  Faculty of Physics, Ludwig-Maximilians-Universit\"at, 
  Scheinerstr. 1, 81679 Munich, Germany
}
\altaffiltext{29}{
  Jet Propulsion Laboratory, California Institute of Technology, 
  4800 Oak Grove Dr., Pasadena, CA 91109, USA
}
\altaffiltext{30}{
  Department of Astronomy, University of Michigan, Ann Arbor, MI 48109, USA
}
\altaffiltext{31}{
  Department of Physics, University of Michigan, Ann Arbor, MI 48109, USA
}
\altaffiltext{32}{
  SLAC National Accelerator Laboratory, Menlo Park, CA 94025, USA
}
\altaffiltext{33}{
  Australian Astronomical Observatory, North Ryde, NSW 2113, Australia
}
\altaffiltext{34}{
  George P. and Cynthia Woods Mitchell Institute for Fundamental Physics and Astronomy, 
  and Department of Physics and Astronomy, 
  Texas A\&M University, College Station, TX 77843,  USA
}
\altaffiltext{35}{
  Instituci\'o Catalana de Recerca i Estudis Avan\c{c}ats, E-08010 Barcelona, Spain
}
\altaffiltext{36}{
  Department of Physics and Astronomy, Pevensey Building, 
  University of Sussex, Brighton, BN1 9QH, UK
}
\altaffiltext{37}{
  Centro de Investigaciones Energ\'eticas, 
  Medioambientales y Tecnol\'ogicas (CIEMAT), Madrid, Spain
}

\begin{abstract}
Host galaxy identification is a crucial step for modern supernova (SN) surveys such as the Dark Energy 
Survey (DES) and the Large Synoptic Survey Telescope (LSST), which will discover SNe by the thousands. 
Spectroscopic resources are limited, so in the absence of real-time SN spectra these surveys must rely on 
host galaxy spectra to obtain accurate redshifts for the Hubble diagram and to improve photometric classification 
of SNe. In addition, SN luminosities are known to correlate with host-galaxy properties. 
Therefore, reliable identification of host galaxies is essential for cosmology and SN science. 
We simulate SN events and their locations within their host galaxies to develop and test methods for 
matching SNe to their hosts. We use both real and simulated galaxy catalog data from the 
Advanced Camera for Surveys General Catalog and MICECATv2.0, respectively.
We also incorporate ``hostless" SNe residing in 
undetected faint hosts into our analysis, with an assumed hostless rate of 5\%. 
Our fully automated algorithm is run on catalog data and matches SNe to their hosts with 91\% accuracy. 
We find that including a machine learning component, run after the initial matching algorithm, improves 
the accuracy (purity) of the matching to 97\% with a 2\% cost in efficiency (true positive rate). 
Although the exact results are dependent on the details of the survey and the galaxy catalogs used, the 
method of identifying host galaxies we outline here can be applied to any transient survey.
\end{abstract}


\section{INTRODUCTION}
\label{Sec:Intro}

A seemingly simple but non-trivial problem that supernova (SN) surveys must confront is how best to match the 
SNe that they discover with their respective host galaxies.  In the absence of spectroscopic or distance 
information about the SNe and the galaxies nearby, matching each SN to its host is a difficult task that is 
impossible to accomplish with complete accuracy.  Although proximity in projected distance 
and spectroscopic redshift agreement between the SN and galaxy are the best indicators we have for positively 
identifying the host, even these indicators are not guaranteed to yield the correct match given that some 
SNe occur in galaxies belonging to pairs, groups, or clusters -- the members of which have similar redshifts.

\begin{figure}[!tb]
\epsscale{1.2}
\plotone{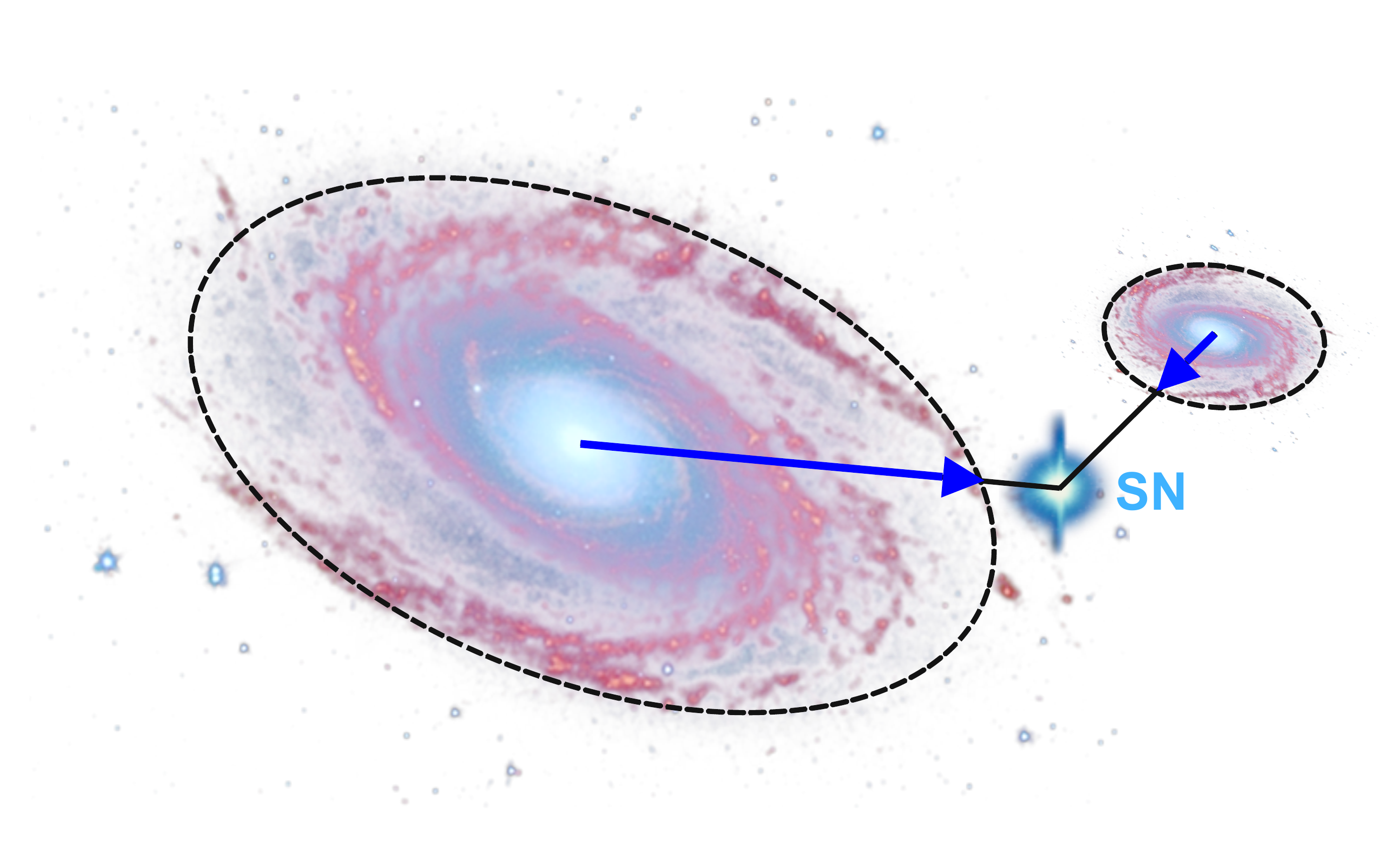}
\caption
{An illustrated example of the problem of host galaxy identification.  The supernova (labeled ``SN") lies in between 
two galaxies. The centroid of the smaller galaxy to the right is closer to the SN in angular separation than the centroid 
of the larger galaxy on the left, but it is possible that the smaller galaxy is a distant background galaxy. 
The blue arrows indicate the light radii of the galaxies (approximated as ellipses) and point toward the SN position.  
This ``directional light radius" (DLR) is discussed in Section~\ref{Sec:HostMatching}.
A real scenario similar to this schematic can be seen in Figure 2 of \citet{daw09}.}
\label{fig:HostGraphic}
\end{figure}

The problem is further compounded by the fact that a small fraction of SNe will occur in dwarf galaxies or 
globular clusters that are too faint to be detected, even in deep stacked images, resulting in so-called ``hostless SNe." 
In particular, the recent new class of SNe known as superluminous SNe \citep{slsne} tend to explode in low-mass 
dwarf galaxies and thus often appear to be hostless upon discovery \citep{bar09,nei11,pap15}. 
There is also evidence that the class of peculiar ``calcium-rich gap" SNe either occur in the outskirts of their 
hosts galaxies (at a projected distance of $> 30\unt{kpc}$) or in low-luminosity hosts \citep{kas12}.
Moreover, truly hostless SNe are possible among intragroup or intracluster stars that have been 
gravitationally stripped from galaxies \citep{gal03,mcg10,san11,gra15}.  In Figure~\ref{fig:HostGraphic} 
we present a schematic illustrating one example of the difficulty in host galaxy identification.

Prior to the era of large SN surveys, the number of SNe discovered was low enough that host galaxies could  
be identified by visual inspection of images. 
With the advent of SN surveys such as the Supernova Legacy Survey (SNLS) and the Sloan Digital Sky Survey-II 
Supernova Survey (SDSS-SNS), came more automated methods. Each of these surveys has thousands of SNe, 
most of which are photometrically identified and thus have no redshift information to aid in host identification. 
For SNLS, \citet{sul06} defined a dimensionless parameter, $R$, that is an elliptical radius derived from 
outputs of SExtractor \citep{sextractor} and computed for every candidate host galaxy. $R$ connects the SN 
position to the galaxy center and is a measure of the SN-host separation normalized by the apparent size of the galaxy.  
For each SN, SNLS selected the galaxy with the smallest value of $R$ as the host, under the condition that $R \leq 5$.
In \citet{sdss:dr}, SDSS-SNS used a method based on \citet{sul06} and defined a quantity termed the directional 
light radius (DLR).  The DLR is the elliptical radius of a galaxy in the direction of the SN in units of arcseconds.  
In Figure~\ref{fig:HostGraphic}, the DLR for each galaxy is represented by the blue arrows.
The dimensionless distance to the SN normalized by DLR is called \ddlr, and this quantity is analogous to $R$. 
For SDSS-SNS, the host matching was performed on all candidate transients by searching within a radius of 30\arcsec\ 
and selecting the galaxy with the minimum \ddlr. There was a nominal restriction which required that the host have 
\ddlr$<4$. However, for a subset of $\sim 100$ SNe the host selected by this algorithm was manually changed after 
visual inspection of images and/or comparisons of redshifts (see Section 8 of \citet{sdss:dr}). This human intervention 
added a bias that cannot be modeled or accurately quantified, and we wish to avoid such issues with host galaxy 
identification in the future, particularly for cases of SNe to be used in cosmological analyses. However, we note that 
visual inspection and human decision are likely necessary for cases of peculiar transients and studies of SN physics. 
The goal of this work is to remove the human subjectivity for cosmologically-useful SNe by using a purely automated 
algorithm, and to use simulations to determine associated biases stemming from incorrect host matches. 

Modern surveys such as the Dark Energy Survey \citep[DES;][]{DES05,ber12} are now discovering SN candidates by the 
thousands. The DES SN Program will discover several thousand SNe Ia over five years, and  
upcoming surveys such as the Large Synoptic Survey Telescope \citep[LSST;][]{LSST} expect to discover hundreds 
of thousands of SNe Ia. Visual inspection of all SN images to identify hosts will be too time-consuming, and a 
determination of the rates of false positives and missed detections cannot be obtained. 
Therefore, a well-defined automated algorithm that can be run on all SN candidates 
is required in order to match SN with their host galaxies and quantify systematic uncertainties.  

Furthermore, the problem of host matching will have a significant impact on cosmology in the near future. 
Given the large number of SNe that will be discovered, acquiring the resources to confirm each spectroscopically 
is an unattainable goal. As a result, we rely predominantly on redshifts obtained from spectra of the host galaxies. 
It is therefore crucial to accurately identify the host galaxy because a misidentified host can result in an incorrect 
redshift assigned to the SN, which will propagate into errors in derived 
cosmological parameters.  Even if the misidentified host has a redshift similar to that of the true host, its host 
properties may be different and thus result in inaccurate corrections for the host-SN luminosity correlation 
\citep{kel10,Sul:10,lam10,gup:11,dan:11,chi13HR,pan14,wol16}.

The method of host galaxy identification that we develop here is applicable to extragalactic transients in general, 
such as gamma-ray bursts, tidal disruption events, and electromagnetic counterparts to gravitational wave sources. 
We are interested in SNe in particular, but classification of a discovered transient often does not 
occur immediately. Therefore, identification of the host galaxy usually comes before classification of the event 
itself, and often aids in the classification process.
In fact, in the absence of a SN spectrum, SN typing relies on a well-sampled light curve and can be 
further improved with a redshift prior from the host galaxy. 
We do not concern ourselves with the details of SN survey detection efficiency for this work. 
We investigate host matching for a range of realistic SN locations, including in galaxies too faint to be detected. 

In this paper, we build upon existing automated algorithms for host galaxy identification such as those implemented in  \citet{sul06} and \citet{sdss:dr}. We go one step further by simulating SN events and placing them in host galaxies to 
test our host matching algorithm's ability to recover the true hosts. We also include a treatment of hostless SNe in our 
analysis and develop a machine learning classifier to compute the probability that our algorithm has matched a SN 
to its correct host.

In Section~\ref{Sec:Methods}, we describe the real and simulated galaxy catalogs from which we draw  
our hosts and also explain the method we use to simulate SN locations.   
In Section~\ref{Sec:HostMatching}, we use the same galaxy catalogs and devise a matching algorithm to pair our SNe to 
their respective host galaxies.  No matching algorithm will be 100\% accurate, so in Section~\ref{Sec:Features} 
we explore features of our host matching results that correlate with correct and wrong matches.
We then examine the benefits of using these features as input into a machine learning classifier (Section~\ref{Sec:ML}), 
trained on simulated data, that returns probabilities of correct matches and helps identify potential cases of 
mismatched host galaxies. In Section~\ref{Sec:Conclusions}, we summarize our findings and outline future work.

\section{METHODS}
\label{Sec:Methods}

We begin by selecting catalogs of galaxies that will serve as hosts for simulated SN locations. 
Our process of simulating SNe (Section~\ref{Sec:SimSN}) and our host-matching algorithm (Section~\ref{Sec:HostMatching}) 
both rely on certain physical characteristics of galaxies, and any galaxy catalog we choose must contain these values. 
Galaxies that are to be selected as SN hosts must have redshifts, preferably spectroscopic, although high-quality 
photometric redshifts (``photo-z's") are also useful.  They must also have morphology or surface brightness profile 
information that will be used to determine the placement of the SNe.  
All galaxies (hosts and galaxies nearby SNe) must have coordinates of their centroids in addition to shape, size, and orientation  
information. We use both a simulated galaxy catalog, where the true properties are known, and also a galaxy catalog  
generated from real data, which is more realistic and more representative of what is available for actual SN surveys.  
We use the simulated (``mock") galaxy catalog to test the algorithm, and then use the real galaxy catalog 
to test if the simulations accurately represent observations. 
In this sense, using both simulations and data serves as a good consistency check. 
Where necessary, we assume a flat $\Lambda$CDM cosmology with $\Omega\ssub{M}=0.3$ and 
$H_0=70\unt{km\ s}^{-1}\unt{Mpc}^{-1}$.

\subsection{Galaxy Catalogs}
\label{Sec:GalCat}

\subsubsection{Simulated Galaxy Catalog}
\label{Sec:MockCat}

For our mock galaxy catalog we use the MICE-Grand Challenge light-cone halo and galaxy catalog release known as 
MICECATv2.0. This catalog was generated by the Marenostrum Institut de Ci{\`e}ncies de 
l'Espai (MICE) collaboration\footnote{\url{http://www.ice.cat/mice}}.  It is complete for DES-like wide-field surveys and 
contains galaxies out to a redshift of 1.4 and down to a magnitude of $i = 24$.  
Beginning with a dark matter halo catalog derived from an $N$-body simulation, the mock galaxy catalog is generated 
from a combination of halo occupation distribution and subhalo abundance matching techniques. 
The catalog was designed to follow local observational constraints, such as the local galaxy luminosity function 
\citep{bla03b,bla05}, galaxy clustering as a function of luminosity and color \citep{zeh11}, and the color-magnitude 
diagram \citep{nyu-vac}.
For details about the input $N$-body simulation and construction of the catalog see \citet{fos15}, \citet{cro15}, and 
\citet{car15}.
The catalog was downloaded via custom query from the CosmoHUB portal\footnote{\url{http://cosmohub.pic.es}}.
We select a $\sim 3$ square-degree region which contains $\sim 300,000$ galaxies. 

The MICECATv2.0 galaxies are modeled as ellipses using a two-component ``bulge-plus-disk" model, with the half-light 
radius of each component given.  It is assumed that the axis ratios for both components are identical. 
Elliptical galaxies are bulge-dominated while spiral galaxies are generally more disk-like.  
Morphological parameters are estimated following \citet{mil13}. MICECATv2.0 uses a color-magnitude selection to 
determine which galaxies are bulge-dominated ({\tt bulge\_fraction} $=1$), following observations from \citet{sch96} and 
\citet{sim02}.  Approximately 15\% of galaxies are bulge-dominated, and the remaining galaxies are disk-dominated and 
have {\tt bulge\_fraction} $<0.4$. 

The galaxies each have a redshift (which includes peculiar velocity), position angle, as well as apparent and absolute 
magnitudes in the DES $grizY$ bands \citep{fla15}. 
Here we work only with the $i$-band magnitude for better comparison with 
our data catalog (Section~\ref{Sec:ACSCat}). There are also galaxy 
properties such as stellar mass, gas-phase metallicity, and star formation rate included in the MICECATv2.0 catalog.
The obvious benefit of the mock catalog is that the true quantities are known. Also, the bulge+disk construction of 
galaxies in MICE provides implicit \sersic\ profile information for all galaxies which is useful for the placement of SNe 
(Section~\ref{Sec:LightProfiles}). 
However, the mocks we use here do not account for instrumental or observational effects that cause problems in real data 
such as the instrument point-spread function (PSF) or deblending detected sources in images. 

\subsubsection{Real Galaxy Catalog}
\label{Sec:ACSCat}

We also use real, high-quality galaxy data from the Advanced Camera for Surveys General Catalog (ACS-GC).  
This is a photometric and morphological database of publicly available data obtained with the Advanced Camera for 
Surveys (ACS) instrument aboard the \textit{Hubble Space Telescope} (\hst) \citep{gri12}. 
The catalog was created using the code \galapagos\ \citep{hau07,hau11}, which incorporates the source detection 
and photometry software SExtractor \citep{sextractor} and the galaxy light profile fitting algorithm GALFIT \citep{galfit}. 

In particular, we use the data from the $\sim 1.8$ square-degree Cosmological Evolutionary Survey 
\citep[COSMOS;][]{cosmos}, which contains approximately 305,000 objects. The COSMOS images were taken with ACS's 
Wide Field Camera (WFC) F814W filter with a scale of $0.05\unt{arcsec\ pixel}^{-1}$ 
and a resolution of $0.09\arcsec$ FWHM. The F814W filter is a broad $i$-band filter spanning the wavelength range 
of roughly $7000 - 9600\unt{\AA}$. The ACS-GC provides 
$\approx 8000$ reasonably secure spectroscopic redshifts from the zCOSMOS redshift survey \citep{zcosmos}. 
In addition, there are $\approx 250,000$ high-quality photo-z's from \citet{ilb09} computed from 30-band photometry 
spanning the UV to mid-IR range. For galaxies with F814W $< 24\unt{mag}$, the median error on photo-z's is 0.02. 
For more about the ACS-GC, see \citet{gri12}.
For galaxies with half-light radii of 0.25\arcsec, the 50\% completeness level is F814W $\simeq 26\unt{mag}$ \citep{cosmos}. 
To approximately match the MICECAT magnitude limit of $i<24$, we impose a brightness limit of F814W $< 24\unt{mag}$ 
which removes 56\% of objects from the ACS-GC. 

Since here we are interested only in a 
catalog of galaxies, we identify compact objects and remove them.  We use the definition of ``compact object" in 
\citet{gri12}, 
i.e., objects with $\mu \leq 18$ or ($\mu \geq 18$ and $r_e \leq 0.03\arcsec$), where $r_e$ is the half-light radius 
determined from GALFIT and $\mu$ is the surface brightness computed from the magnitude and ellipse area. 
Excluding these removes an additional 9\% of objects.  
We have confirmed that after removing compact objects and requiring F814W $< 24\unt{mag}$ the average galaxy density 
(number per square arcmin) agrees with MICECATv2.0, with some difference expected due to differences between the 
DES $i$ filter ($\approx 7000 - 8500\unt{\AA}$) and the \hst\ F814W filter ($\approx 7000 - 9600\unt{\AA}$) and the fact that 
both catalogs are cut at magnitude 24.

\subsection{Simulating Supernovae in Host Galaxies}
\label{Sec:SimSN}
\citet{Kel08} studied the distribution of SNe within their host galaxies and found that SNe Ia as well as SNe II and 
SNe Ib track their host galaxy's light.  Therefore, for the purpose of this study, it seems reasonable to use the  
surface brightness profile of a galaxy to determine the placement of a simulated SN location within it. 
In addition, since the probability of a SN occurring in a galaxy is roughly proportional to the mass of the  
galaxy \citep{sul06,smi12}, which is in turn proportional to the luminosity, when selecting host galaxies we weight by the 
galaxy luminosity. We describe this process in more detail below. 

\subsubsection{Host Galaxy Light Profiles}
\label{Sec:LightProfiles}

We use the supernova analysis software package \snana\footnote{\url{http://das.sdss2.org/ge/sample/sdsssn/SNANA-PUBLIC/}}
 \citep{kes09b} to determine the placement of simulated SN locations onto host galaxies.  
This software was used to place simulated SNe (a.k.a. ``fakes") onto real galaxies for monitoring of  
the difference imaging pipeline and the detection efficiency of the DES Supernova Program \citep{des:diffim}.
The placement of SNe requires an input galaxy catalog that serves as a ``host library" and contains 
information such as galaxy positions, redshifts, magnitudes, orientations, shapes, sizes, and light profile parameters. 

For each simulated SN, a random host galaxy is selected from the input host library, under the condition that the redshift of 
the galaxy matches the redshift of the SN to within 0.001. For the subset of galaxies that satisfy this redshift agreement 
criterion we then weight the galaxies by their luminosity, assuming a simplistic linear probability function such that galaxies 
with higher luminosity are preferred over those with lower luminosity. For MICECAT the absolute magnitudes are provided 
and so we convert the DES $i$-band absolute magnitude into a luminosity and use this as the weight. For ACS-GC, no 
absolute magnitudes are provided and so instead we compute a pseudo-absolute magnitude defined as the apparent 
magnitude in the F814W filter minus the distance modulus (calculated from the galaxy redshift and our assumed cosmology). 
We ignore $K$-corrections which are typically $\lesssim 1$ mag and increase with redshift on average. 
This pseudo-absolute magnitude is then converted into a luminosity which is used as the weight. 
Once a suitable host is selected, the exact coordinates of the SN are chosen by randomly sampling from the host's light 
profile so that the probability of the SN being at a particular location relative to the host galaxy center is weighted 
by the host's surface brightness. The actual redshifts and coordinates of the potential host galaxies in the catalog 
are used in determining the placement of SNe. 

Galaxy brightness profiles are often described by a \sersic\ profile \citep{ser63}, 
which gives brightness, $I$, as a function of distance from the galactic center, $r$:
\begin{equation}
I(r) = I_0 \exp{\LR{-b_n \lr{\frac{r}{r_e}}^{1/n} }},
\label{eqn:Sersic}
\end{equation}
where $r_e$ is the half-light radius, $n$ is the \sersic\ index, and $b_n$ is a constant that depends on $n$.  
For details on \sersic\ profiles see \citet{cio91} and \citet{gd05}. 
A profile with $n=4$ is known as a de Vaucouleurs profile \citep{deV48} and is generally a good fit to 
elliptical galaxies.  A profile with $n=1$ is an exponential profile, which is a good description of disk galaxies.  
Galaxies with large values of $n$ are more centrally-concentrated, but also contain more light at large $r$, in the 
wings of the distribution. 

When creating the host library for the MICECAT galaxies, we assume that the bulge component 
of the MICE mock galaxies has a de Vaucouleurs profile while the disk component has an exponential profile. 
The half-light radii for each component are given by the catalog parameters {\tt bulge\_length} and {\tt disk\_length}.
The {\tt bulge\_fraction} provides the weight given to the bulge component, and \snana\ is 
able to construct weighted sums of \sersic\ profiles and thus the total light profile for each galaxy in the host library. 
The axis ratio and position angle together with the light profile of each host galaxy are used to simulate 
the SN position.

For the ACS-GC galaxies, GALFIT was used by \citet{gri12} to simultaneously fit a half-light radius $r_e$ and a 
\sersic\ index in the range $0.2 \leq n \leq 8.0$. 
We use this single fitted \sersic\ profile to reconstruct the light profile in \snana. This light profile along with 
the axis ratio and position angle determined by GALFIT are used to simulate the SN position.
To help ensure that our ACS-GC host galaxies are truly galaxies and that they have well-measured light profile 
parameters for placing simulated SNe, we create an ACS-GC host library by imposing the following selection criteria 
on sources. In parentheses we list the cumulative fraction of the total ACS-GC sample remaining after each additional 
criterion is imposed. We require each host    
\begin{enumerate}
\item Have a F814W magnitude $< 24$ (43.6\%)
\item Not be a compact source, where ``compact source" is defined as in \citet{gri12} 
and Section~\ref{Sec:ACSCat} (37.8\%)
\item Have a redshift in the catalog (36.6\%)
\item Have errors on the GALFIT \sersic\ parameters $r_e$ and $n$ that are $<15\%$ and 
have values of $r_e$ and $n$ not identically equal the maximum allowed values 
(max$\Lr{r_e} = 37.5\arcsec$, max$\Lr{n} = 8.0$), since those cases are often indicative of failures in the fits (30.8\%)
\end{enumerate}
This leaves us with $\approx 94,000$ galaxies as potential hosts. 
These requirements are intended to maintain the balance between reliability of the host-galaxy parameters 
and the bias against faint galaxies whose measured properties are more uncertain. While the selection criteria listed 
above will still allow some fraction of galaxies with faulty GALFIT parameters to serve as hosts, we find that only 1\% of 
our selected host galaxies have extreme values of $r_e > 5\arcsec$. 
We have run tests where we modified the values of the \sersic\ indices in the host library and found that the 
effect of the \sersic\ index is subdominant to the effect of size of the half-light radius when it comes to the simulated 
SN-host separation. 

\subsubsection{Redshift Distribution}
For the purposes of testing algorithms to identify the host galaxy, the SN coordinates are the only relevant SN quantity. 
In order to have a realistic redshift distribution similar to that of an actual SN survey, we simulate  
SNe Ia with the observing conditions and detection efficiency of the DES SN Program. 
We assume the SN Ia rate from \citet{dil08} (i.e., $(2.6\E{-5}) \times (1+z)^{1.5}\unt{SNe\ Mpc}^{-3}\unt{yr}^{-1}$), 
which was also assumed in \citet{ber12}. 
We simulate SNe in the range $0.08 < z < 1.4$ as these are the redshift limits of MICECATv2.0. 
\snana\ generates each redshift from a random comoving volume element weighted by the SN Ia volumetric rate and 
selects a host from the host library that matches the redshift with a tolerance which we have set to 0.001.
Since there is less volume at lower redshifts and we intend to simulate many SNe, 
we allow for individual galaxies in the host library to host more than one SN. 
This does not pose a problem for this study since each SN is drawn from a different random number which 
is used to place it. As a result, a particular galaxy may be a host for multiple SNe, 
but each SN will have an independent random orientation with respect to that host. 

\begin{figure*}[!tb]
\epsscale{1.0}
\plotone{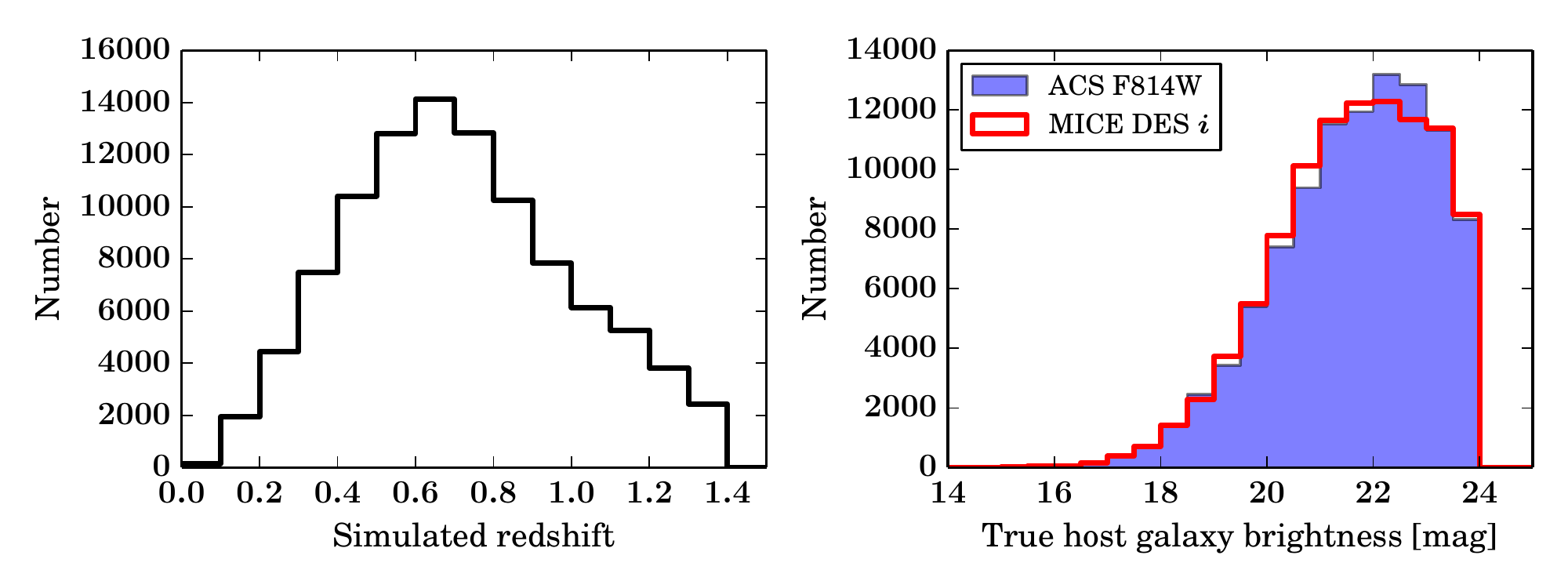}
\caption{\emph{Left}: Redshift distribution for the 100K SNe simulated on MICE and ACS-GC galaxies. 
By construction, the redshift distributions for MICE and ACS-GC are nearly identical. 
\emph{Right}: The host galaxy magnitude distribution for these SNe. The ACS-GC host magnitudes 
measured in the F814W filter by SExtractor (MAG\_BEST) are shown in filled blue; the MICE host magnitudes 
in the DES $i$ filter are shown in red.}
\label{fig:zDist}
\end{figure*}

We simulate SN Ia light curves using the SALT2 model \citep{guy07} 
and the measured SN cadence and observing conditions of the first 2.5 years of the DES SN survey. 
To sculpt the redshift distribution we apply the DES detection efficiency as a function of $S/N$  
derived from DES SN Year 1 data \citep{die14} and impose the DES transient trigger criterion of 2 detections 
in any filter, occurring on different nights. 
We simulate 100,000 SNe each on the MICECAT and ACS-GC galaxies, using their respective host libraries 
and each satisfying the DES trigger criterion.  The resulting redshift distribution (which is the same for both MICECAT 
and ACS-GC by construction) as well as the 
magnitude distribution of the hosts is shown in Figure~\ref{fig:zDist}.

Here we have ignored Milky Way extinction and Poisson noise from the host galaxy when 
simulating our SNe and computing $S/N$. 
We emphasize that the goal of this simulation is purely to obtain a redshift distribution that is somewhat 
realistic, and the details of the generated SNe Ia and their light curves are not relevant here.  A more detailed simulation 
(including galaxies measured by DES, galactic extinction, fits to light curves) is planned for a future paper.

\subsubsection{Comparison with SN Data}

\begin{figure}[!tb]
\epsscale{1.2}
\plotone{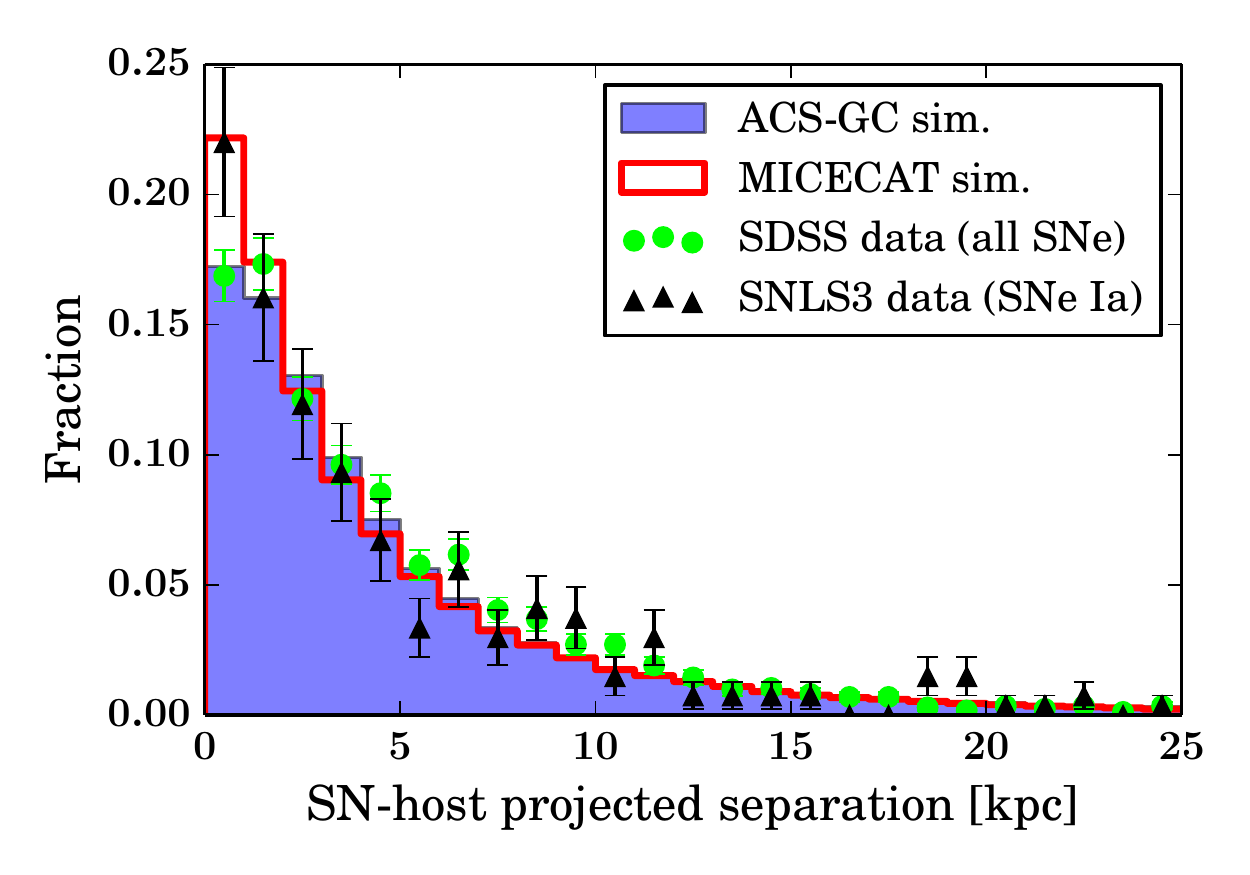}
\caption{Distribution of the SN-host projected separation for our SN simulations using both  
ACS-GC (filled blue histogram) and MICECATv2.0 (red histogram) galaxy catalogs (100,000 SNe each).  
For comparison with data we also show 1737 SNe from SDSS-SNS (green circles) and 
268 SNe from SNLS3 (black triangles).}
\label{fig:kpcDist}
\end{figure}

We find that our host galaxy (pseudo-)absolute magnitude distributions appear roughly consistent with the SN Ia 
host galaxy SDSS $i$-band absolute magnitude distribution derived from SDSS data in \citet{yf10}.  
To check that we are placing SNe at reasonable separation distances from their hosts given the MICECAT and 
ACS-GC host libraries, we plot the distribution of SN-host separations and compare to actual SN survey data. 
Rather than comparing the SN-host angular separations, we compare projected SN-host separation distance, 
in units of kpc, to account for the differences in redshift distributions between different surveys.
This quantity is shown in Figure~\ref{fig:kpcDist}, where we overplot data for the SNe from the SDSS-SNS 
and SNLS3 that have identified host galaxies and compare them with our simulated distributions.  
The SDSS-SNS data includes 1737 spectroscopically-confirmed or photometrically-classified SNe (with 
host-galaxy spectroscopic redshifts) of all SN types with hosts from \citet{sdss:dr}, 
while the SNLS3 data includes only the 268 spectroscopically-confirmed SNe Ia with hosts published in \citet{guy10}. 
In general, our simulated SNe show very good agreement with data, indicating that our methods are sensible. 

The two datasets (SDSS and SNLS) agree fairly well within errors, although SDSS seems to be less efficient than SNLS at 
detecting SNe near the core of the galaxy, as seen in the first bin in Figure~\ref{fig:kpcDist}. 
This difference might be partly explained by the SDSS SN spectroscopic follow-up strategy. 
We have confirmed in the data that the spectroscopically-confirmed SDSS SNe are biased against SNe near galactic 
cores when compared to the photometrically-typed SNe (whose redshifts were obtained from host-galaxy spectra taken 
after the SNe had faded away). Since SDSS was a lower-redshift survey compared to SNLS, contamination from 
bright, relatively nearby hosts likely prevented SDSS from obtaining some SNe spectra. 

\begin{figure}[!tb]
\epsscale{1.1}
\plotone{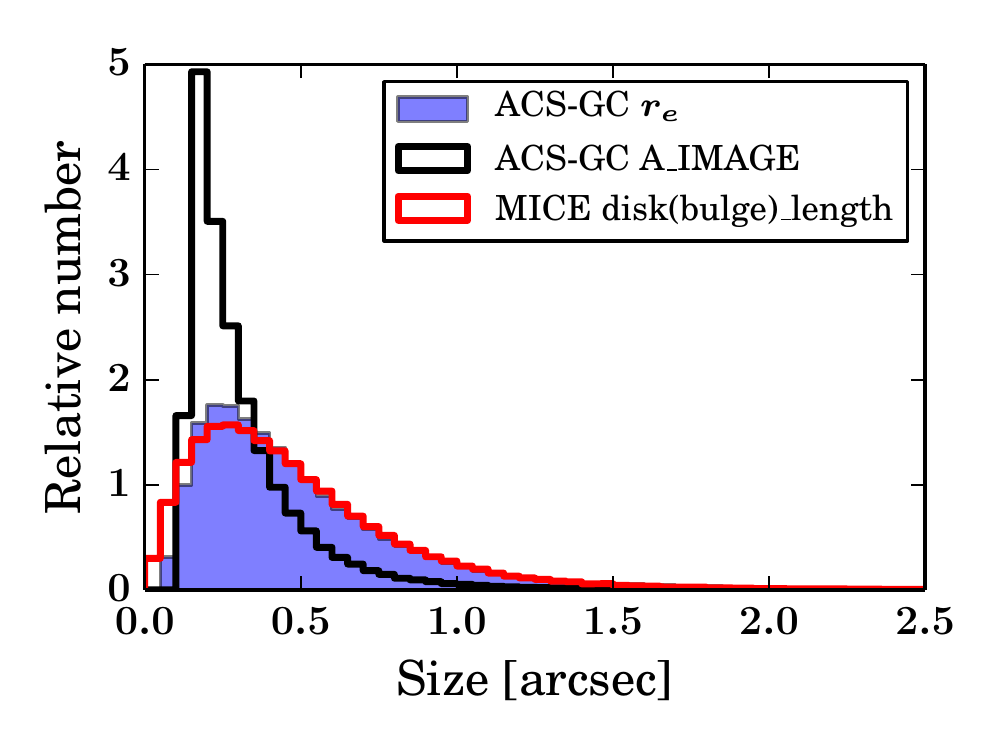}
\caption{Comparison of sizes for galaxies in the MICECATv2.0 and ACS-GC COSMOS host libraries. 
For MICE, the galaxy size plotted is the {\tt bulge\_length} for bulge-dominated galaxies and the 
{\tt disk\_length}, otherwise.}
\label{fig:GalSizes}
\end{figure}

The distribution of simulated SN-host separation on MICECAT galaxies and ACS-GC galaxies also agree quite well with each 
other. This is not surprising given that the distribution of galaxy sizes are very similar between the two catalogs. This can be  
seen in Figure~\ref{fig:GalSizes} when comparing ACS-GC $r_e$ sizes (blue filled histogram) to the MICECAT sizes (red 
open histogram). For the MICECAT sizes we plot {\tt bulge\_length} for bulge-dominated galaxies and the 
{\tt disk\_length}, otherwise.  The similarity in the  ACS-GC $r_e$ and MICECAT size distributions makes sense since 
both are half-light radii derived from \hst\ data\footnote{MICECAT sizes are simulated from relations 
derived from \hst\ data \citep{mil13, sim02}}.
However, there is an excess of SNe at low SN-host separations in MICECAT compared to 
ACS-GC (first two bins in Figure~\ref{fig:kpcDist}).  This is likely due to the excess of  
small galaxies seen in MICE in Figure~\ref{fig:GalSizes}. ACS-GC sizes are limited by the PSF of the \hst\ images 
($0.09$ arcsec), while the minimum size of MICECAT galaxies is $10^{-4}$ arcsec. Such small galaxies in MICECAT would
go unresolved in ACS-GC and thus would appear larger. 

For ACS-GC, we also show in Figure~\ref{fig:GalSizes} the {\tt A\_IMAGE} value from SExtractor (black open histogram), 
which is used to perform the host matching (Section~\ref{Sec:DLR}). {\tt A\_IMAGE} is a measure of size derived from the 
second moments of the light distribution in the raw images; unlike $r_e$, it is not derived from fitting a model. 
For galaxies that are well-measured with GALFIT there is a tight linear relationship between $r_e$ and {\tt A\_IMAGE}.

\section{HOST MATCHING ALGORITHM}
\label{Sec:HostMatching}
\subsection{Directional Light Radius (DLR) Method}
\label{Sec:DLR}

We employ the DLR host matching method used for the final data release of the SDSS-SNS and described in \citet{sdss:dr}. 
As mentioned in the Introduction, this method is similar to that developed by \citet{sul06} for SNLS.   
Explicitly, the distance from a SN to a nearby galaxy, normalized by the galaxy's DLR is termed \ddlr\ and is defined as 
\begin{equation}
d_{\mathrm{DLR}} = \frac{\text{SN--galaxy angular separation (arcsec)}}{\text{DLR (arcsec)}}
\end{equation}

Our method assumes that galaxies in images are elliptical in shape and can be described by a semi-major axis $A$ and a 
semi-minor axis $B$.  In addition, the galaxy position angle $\phi$ is the orientation of $A$ relative to a fixed coordinate 
axis on the sky.  Given these quantities for each galaxy along with the coordinates of the SN, we can compute \ddlr.  
When matching a SN to its host, we first begin by searching for all galaxies within $30\arcsec$ of the SN position. 
We compute \ddlr\ for each of these galaxies and order them by increasing \ddlr.  The nearest galaxy in \ddlr-space (i.e., 
the galaxy with the minimum \ddlr) is designated as the host. Based on our simulations, 0.05\% of MICE SNe and 0.6\% of 
ACS-GC SNe are actually located $>30\arcsec$ from the center of their hosts, and we remove these SNe from our sample. 
This rate is higher in ACS-GC because despite our fairly strict host library selection criteria (Section~\ref{Sec:LightProfiles}), 
some galaxies still have poorly-fit light profiles resulting in \sersic\ $n$ and $r_e$ values that are too large, which in turn results 
in SNe being simulated at extreme separation distances from their hosts. 
However, it is worthwhile to note that it is possible that some small fraction of low-redshift SN will be located at large angular 
separations from their hosts. 

We emphasize that DLR is a survey-dependent quantity 
as it relies on measures of $A$ and $B$ which are themselves survey-dependent. 
For example, measurements of the shape and size of galaxies depend on the image filters and PSFs.  
Furthermore, the algorithm used to make these measurements may differ between surveys as well.  
For MICECAT, each galaxy has only a {\tt disk\_length} and a {\tt bulge\_length}. Therefore, when matching SNe to galaxies, 
we assume that a galaxy has a semi-major axis equal to {\tt bulge\_length} if {\tt bulge\_fraction} $=1$ and equal to 
{\tt disk\_length}, otherwise (bulge fractions that are not identically unity are all $<0.4$).  
This semi-major axis is plotted for the MICECAT sizes in Figure~\ref{fig:GalSizes}. 
For ACS-GC, we use the fitted GALFIT position angle, axis ratio, \sersic\ index $n$, and size scale $r_e$ 
in the host library when placing the SNe, but use the measured SExtractor parameters 
{\tt A\_IMAGE}, {\tt B\_IMAGE}, and {\tt THETA\_IMAGE} when computing DLR and performing the matching, 
since these types of parameters are more readily available in a real survey catalog. 
We find that matching using $r_e$ to compute DLR for all ACS-GC galaxies within the search radius 
results in a greatly reduced matching accuracy. This is due to the fact that, in the absence of a quality cut on GALFIT 
parameters when performing the matching, some of the fainter galaxies nearby the SN can have unreliable values of $r_e$. 
These poorly-fit galaxies tend to have $r_e$ values that are biased to be too large, which results in their DLR separation 
from the SN being very small. This causes them to be preferentially selected (incorrectly) as the host since the 
matching criterion is minimum DLR separation, leading to a reduced matching accuracy. By contrast, the SExtractor 
parameters we use are not fits to any model and are more robust size estimates in cases of faint or blended galaxies.

\subsection{Magnitude Limits \& Hostless SNe}
\label{Sec:MagLim}

A magnitude-limited SN survey will detect some fraction of SNe in low-luminosity galaxies that fall below 
this magnitude limit. We wish to understand the effect of such hostless SNe on host matching.  As an example, 
for the real SN data used in Figure~\ref{fig:kpcDist}, $\approx 6\%$ of the SNLS SNe and $\approx 4\%$ of the 
SDSS SNe were excluded from the figure because they had no identified host. For SNLS, ``hostless" was defined as 
having no galaxies within $5R$ \citep{sul06}, and for SDSS the nominal definition was having no galaxies within 
4 \ddlr, but with some manual corrections based on visual inspection and redshift agreement \citep{sdss:dr}. 
The problem with these definitions is that they do not distinguish between cases where the true 
host is detected but simply too far away (above the distance threshold) and cases where the true host is 
too faint to be detected. In the former case the host can be recovered by increasing the (arbitrary) distance threshold 
for matching. The latter case is more worrisome since some of the time the true host will not be detected and yet some 
other (brighter) galaxy could fall within the distance threshold, resulting in a misidentified host. Therefore, it is this 
latter case that we focus our attention on for this paper. 
Here we select a fiducial hostless rate of 5\% and na\"ively assume that these SNe are hostless because the 
true host is fainter than the magnitude limit. Our definition of ``hostless" here therefore differs from the definitions of 
SDSS and SNLS, where ``hostless" could simply mean the true host lies beyond a certain 
distance threshold.  Also our definition does not account for the possibility of SNe occurring outside of galaxies, 
within the intragroup or intracluster medium. However, we believe that our treatment of hostless galaxies is  
sufficient for the illustrative purpose of this study. 

\begin{figure}[!tb]
\epsscale{1.1}
\plotone{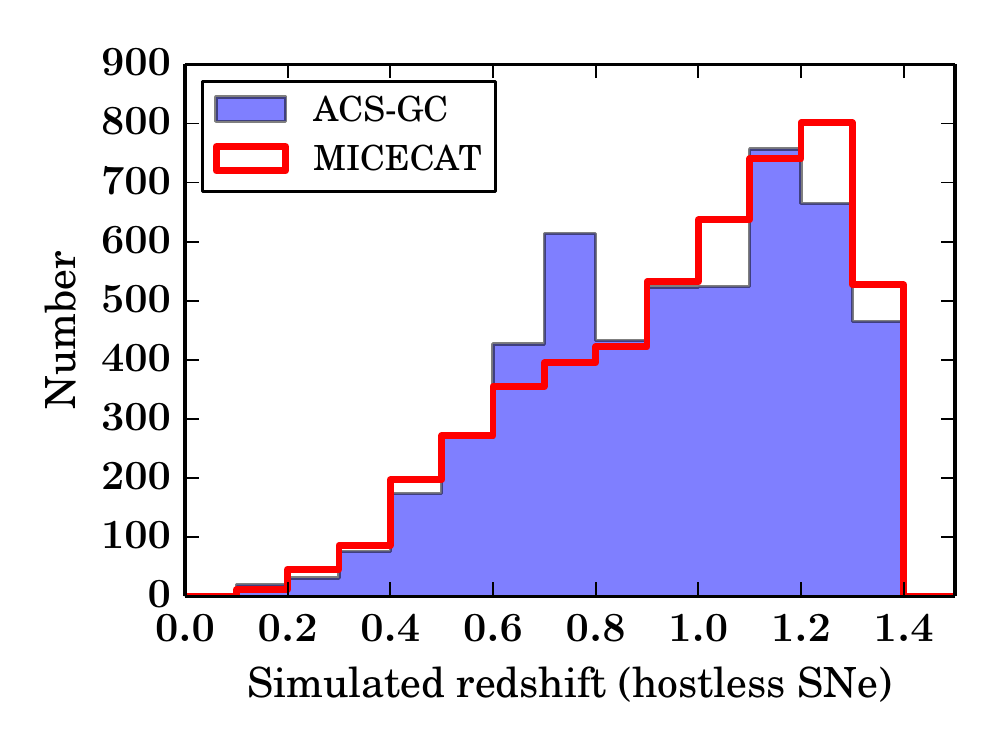}
\caption{Redshift distribution of the 5\% of SNe taken to be hostless. As the hostless sample was created by 
imposing a magnitude limit, the number increases with redshift.}
\label{fig:hostless_z}
\end{figure}

To create our hostless sample, we impose a magnitude limit on our galaxy catalogs when performing 
the matching such that 5\% of our simulated SNe are hosted by galaxies with brightnesses below this limit. 
These limits are $i\ssub{lim}=23.67$ for MICECAT and F814W$\ssub{lim}=23.68$ for ACS-GC. 
Thus, when running our host-matching algorithm we first remove galaxies fainter than the magnitude limit, thereby 
creating hostless SNe comprising 5\% of our sample for which we know the hosts will be incorrectly matched to 
galaxies brighter than the true host. Fixing the hostless rate to 5\% for both galaxy catalogs allows us to better 
compare the matching accuracies. As seen in Figure~\ref{fig:hostless_z}, our number of hostless SNe increases with redshift, 
which is expected since galaxies at higher redshift are generally fainter. There is an indication of a similar trend for the 
hostless SNe in SNLS, though the statistics are low. For SDSS, the redshift distribution for hostless SNe is flatter, but 
the redshift range of SDSS is roughly half the range of SNLS. Also, the SDSS sample includes photometrically-classified 
SNe with host galaxy redshifts, which by construction cannot be hostless. 

Our study is limited by the magnitude depth of our chosen galaxy catalogs, both simulated and real. 
Current and future surveys will eventually surpass these in depth, revealing even fainter galaxies.  In fact, even 
our DES-like MICE catalog is only complete out to $i = 24$, which is the estimated five-year depth of the 
DES wide-field survey.  However, the DES SN fields are observed more frequently and attain a one-season co-add 
$5\sigma$ limiting magnitude of $\sim 26$ for point sources in the shallow fields and $\sim 27$ in the deep fields, 
which will increase to $\sim 0.85$ mag deeper when the full five seasons are co-added \citep{ber12}. 
We also point out that the true rate of hostless SNe in any survey depends on the specifics of the survey,  
the SN type, and the host galaxy luminosity functions (LFs) for those respective types, among other things. 
For the purpose of this analysis we believe a 5\% hostless rate to be a reasonable assumption. 
In a future paper, we intend to focus specifically on matching the hosts of SNe Ia, and we plan to implement prior 
knowledge of the SN Ia host galaxy LFs into our simulations. 

\subsection{Results \& Performance}
\label{Sec:MatchResults}

Our main method of host matching is the DLR method described in the previous section.  A summary of the host 
matching results for both MICECAT and ACS-GC is given in Table~\ref{tab:matchres}. We also match 
based on nearest angular separation since this is the simplest and computationally easiest method. 
This method agrees with the DLR method 91\% of the time for MICECAT and 95\% of the time for ACS-GC. 
However, the DLR method slightly outperforms the angular separation method for both catalogs. 
We find that when using MICECAT, the DLR matching accuracy is 90.11\% and the nearest separation 
matching accuracy is 88.35\%. When using ACS-GC, the DLR matching accuracy is 92.21\% 
and the nearest separation matching accuracy is 90.62\%.  Recall that 5\% of the mismatch is due to 
hostless SNe which get matched to galaxies brighter than their true hosts. 
For MICECAT, the 2nd-nearest and 3rd-nearest galaxies in DLR are the true 
host 4\% and 0.6\% of the time, respectively. For ACS-GC, these values are 2\% and 0.5\%. In cases where 
the nearest DLR galaxy is not the correct host, the nearest galaxy in angular separation is the correct host 
2\% of the time in MICECAT and 0.5\% of the time in ACS-GC. 

In order to understand why the overall DLR matching accuracy is higher for ACS-GC than for MICECAT galaxies 
by $2.11 \pm 0.13$\% we return to Figures~\ref{fig:kpcDist} and \ref{fig:GalSizes}. 
The simulated SN-host separations and true host galaxy sizes are not different enough to 
account for this difference in matching accuracy between ACS-GC and MICECAT. 
Another factor that might be responsible is the galaxy spatial distributions and clustering properties 
of the two catalogs. A related issue is the detection and deblending of galaxies in ACS-GC. 
We investigate differences in the galaxy clustering of the two catalogs in the Appendix. The main result is that at small 
angular separations ($< 2\arcsec$), MICECAT exhibits a much higher number of galaxy pairs relative to ACS-GC.
In addition, it is 
common for MICECAT galaxy pairs at this separation to overlap or occlude each other. Whether or not this clustering 
accurately represents true galaxy dynamics is unclear. However, if a high degree of small-scale clustering does exist,
such galaxy pairs in real data would be difficult to separate or even impossible to see if completely occluded and may 
be identified as a single galaxy in the catalog. This would explain in part the decreased galaxy number density at 
small scales in ACS-GC and thus the slightly higher overall matching accuracy when compared to MICECAT. 
Looking specifically at our true host galaxies, we find that while the mismatch rate for true hosts with neighbors within 
2\arcsec\ is similar for both MICECAT and ACS-GC, the occurrence of true hosts with neighbors this close is much higher 
for MICECAT (22\% of all hosts) than for ACS-GC (only 4\% of all hosts).

\begin{deluxetable*}{lcc}[htb]
\tablecaption{Summary of Host Matching Results \label{tab:matchres}}
\tablewidth{0pt}
\tablehead{
\colhead{} & \multicolumn{2}{c}{\textit{Galaxy Catalog}}\\
\colhead{} & \colhead{MICECATv2.0} & \colhead{ACS-GC COSMOS} 
}
\startdata 
Accuracy, nearest separation$^a$ & $88.35\pm 0.10$\% & $90.62\pm 0.09$\% \\
Accuracy, DLR method$^a$ & $90.11\pm 0.09$\% & $92.21\pm 0.09$\% \\
Accuracy (purity), DLR cut$^b$ & $94.45\pm 0.09$\% & $97.29\pm 0.09$\% \\
Accuracy (purity), ML cut$^b$ & $96.19\pm 0.19$\% & $97.71\pm 0.16$\% \\
\enddata
\tablenotetext{a}{Purity at 100\% efficiency}
\tablenotetext{b}{Purity at 98\% efficiency; objects removed by cut}
\tablecomments{Accuracies include hostless SNe. The accuracy after ML is based on simulations of 10K SNe; 
the other accuracies are derived from an independent set of 100K SNe.}
\end{deluxetable*}

\begin{figure*}[tb]
\epsscale{0.9}
\plotone{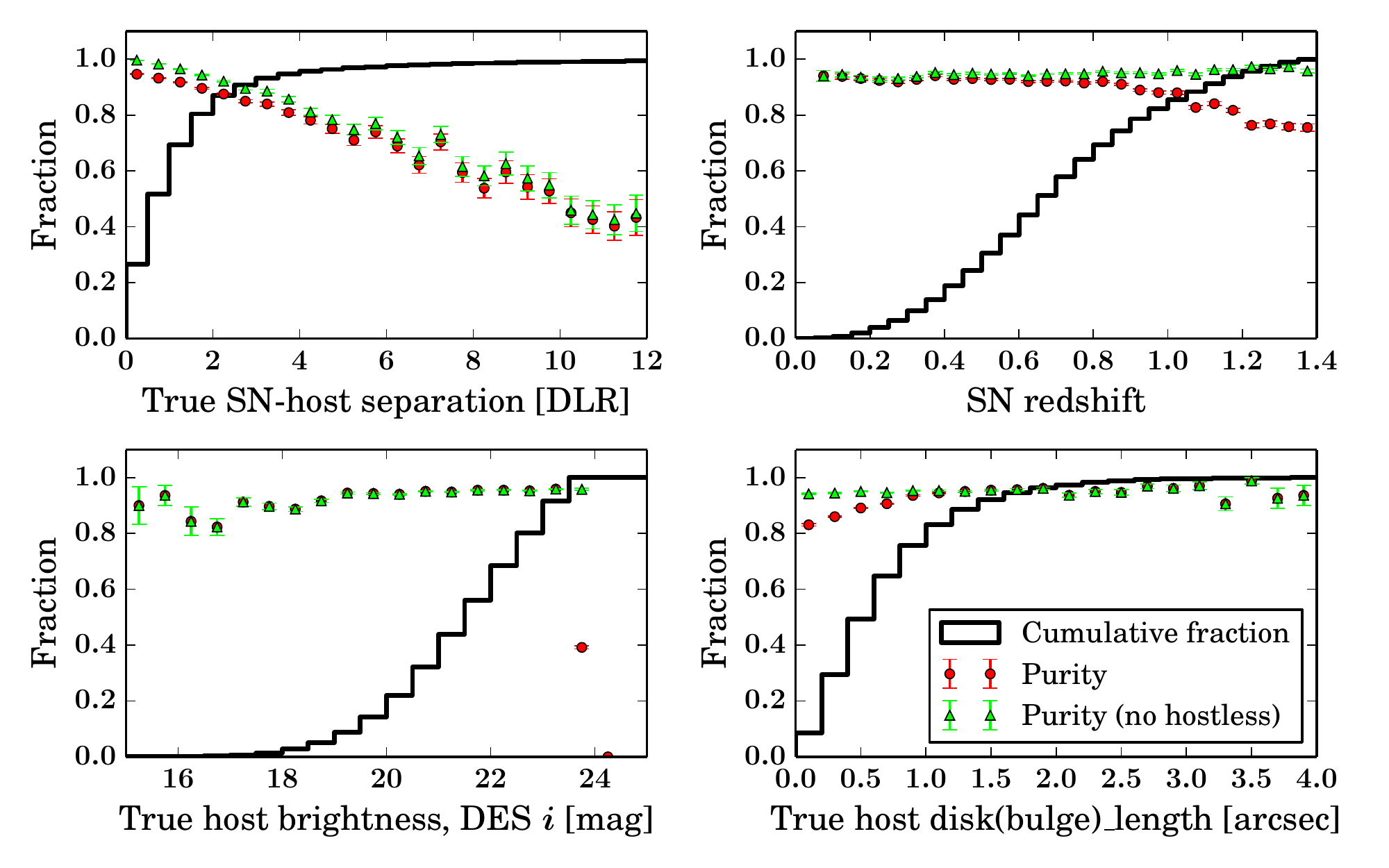}
\caption{DLR matching accuracy (purity) as a function of true SN-host separation, redshift, true host brightness, and 
true host size for the SNe simulated on MICECATv2.0 galaxies. The purity is given for all SNe (red circles) and also for 
the sample that excludes hostless SNe (green triangles). The black histogram is the cumulative fraction for all 
simulated SNe.}
\label{fig:MICE_EffPur}
\end{figure*}

\begin{figure*}[tb]
\epsscale{0.9}
\plotone{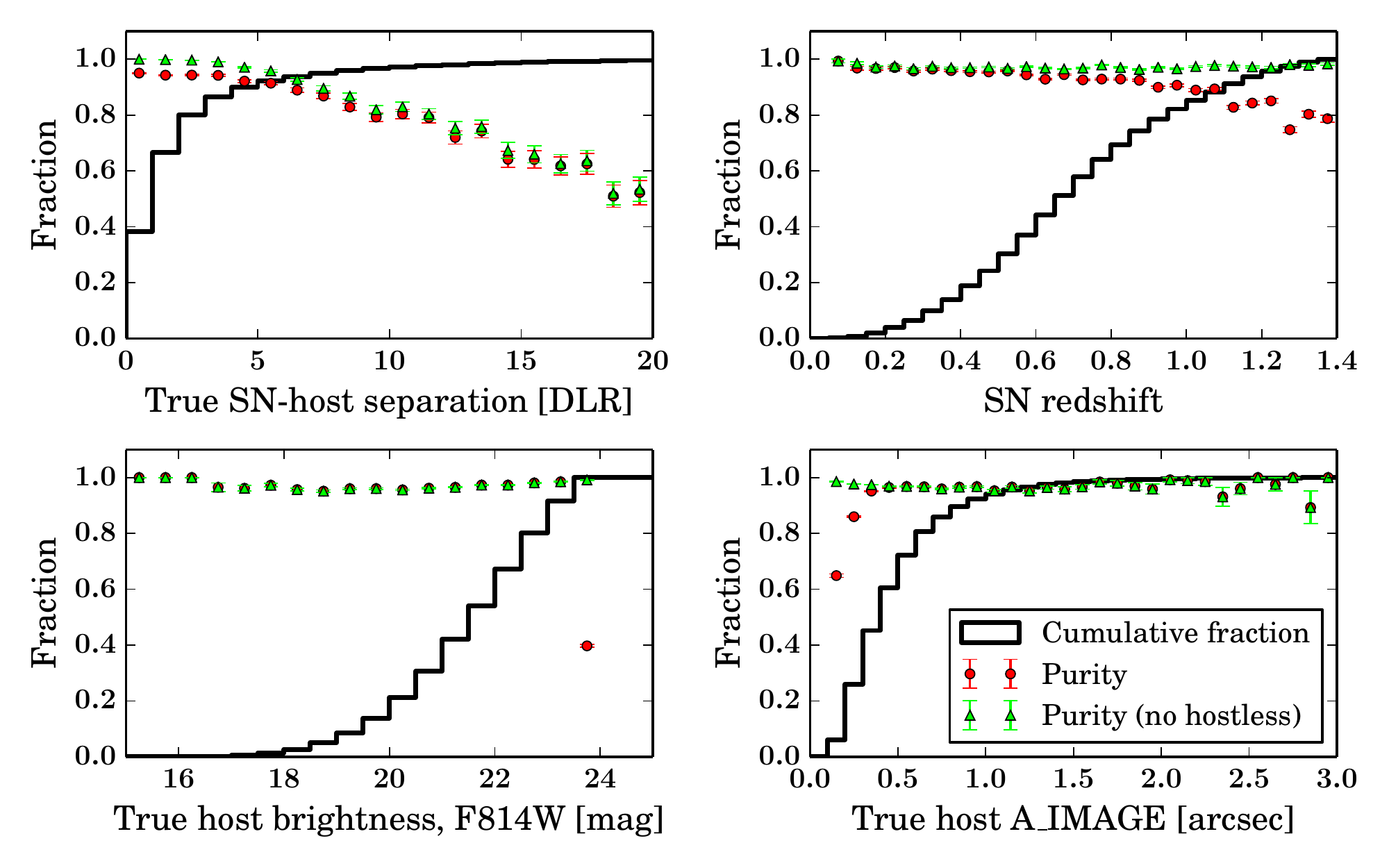}
\caption{Same as Figure~\ref{fig:MICE_EffPur} but for the SNe simulated on ACS-GC galaxies.}
\label{fig:ACSGC_EffPur}
\end{figure*}

In Figures~\ref{fig:MICE_EffPur} and \ref{fig:ACSGC_EffPur} we plot the matching accuracy (purity) as a function of 
SN-true host separation, SN redshift, true host magnitude, and true host size for both the MICE 
and ACS-GC cases, respectively. We show both the purity for the entire sample (red circles) and also for the sample with 
hostless SNe removed (green triangles) in order to better see the effect of the hostless SNe.  
We also show the cumulative fractions for all simulated SNe as the black histograms. 
The matching accuracy is highly sensitive to the separation from the true host, 
as one would expect since SNe that are farther away from their hosts have a higher probability of 
being matched to another nearby galaxy. Note that the exact DLR values cannot be directly compared 
between MICECAT and ACS-GC, as they are computed using different measures of galaxy size. 
The hostless SNe reduce the purity at smaller values of true host separation since the true hosts are faint and 
generally small, which results in the SNe often being simulated near the host center. 

\begin{figure}[!tb]
\epsscale{1.2}
\plotone{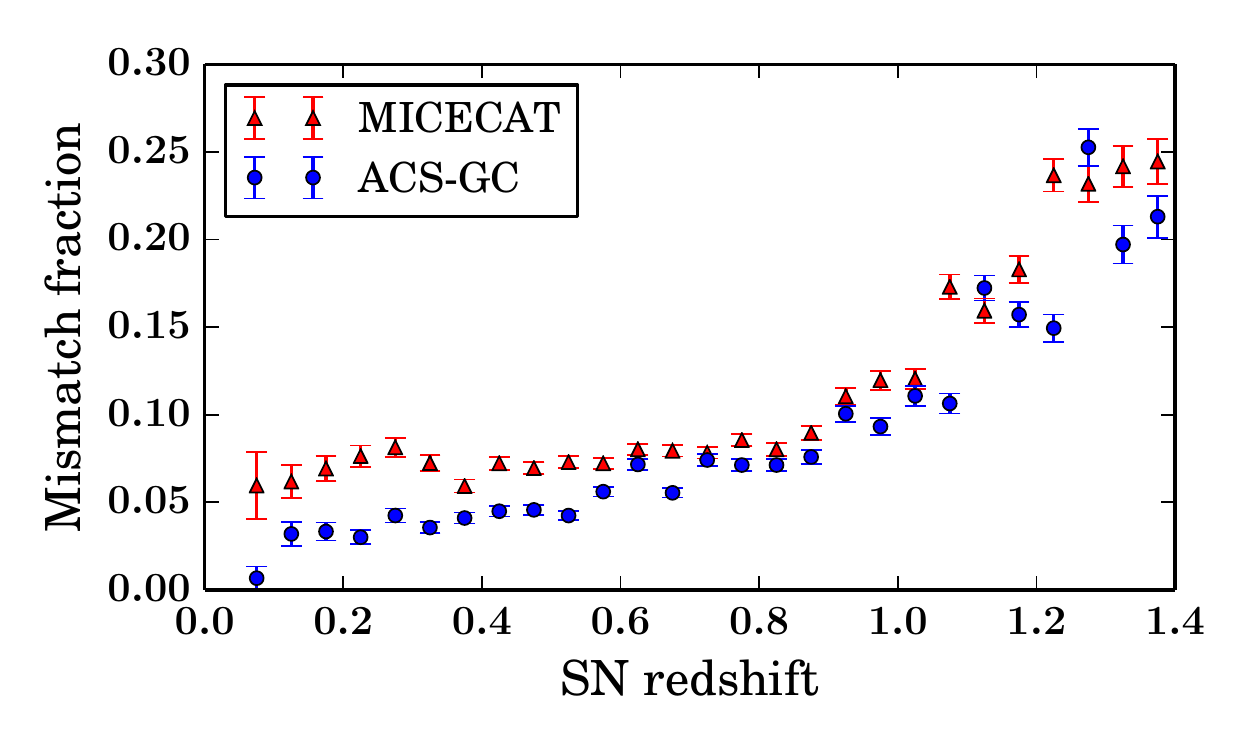}
\caption{The host-galaxy mismatch fraction as a function of redshift for both MICECAT and ACS-GC.}
\label{fig:MismatchVz}
\end{figure}

The purity as a function of redshift is constant for $z \lesssim 0.6$, but begins to drop significantly at higher redshifts 
due to an increase in the rate of hostless SNe which reside in the faintest galaxies. 
A plot of the mismatch fraction ($= 1-$ purity) versus redshift is shown in Figure~\ref{fig:MismatchVz} with 
the results for MICECAT and ACS-GC overlaid for better comparison. The trend with redshift is similar for both catalogs, 
with MICECAT offset from ACS-GC due to the overall lower matching accuracy of MICECAT. 
The purity (and mismatch fraction) is fairly constant at all redshifts for both catalogs once the hostless SNe are removed.  

For both MICECAT and ACS-GC, the matching purity is fairly insensitive to the true host galaxy brightness except for 
the faintest hosts where the purity drops precipitously, as expected due the magnitude limit we impose for our hostless 
SNe (Section~\ref{Sec:MagLim}). In both catalogs, the matching purity is lower for the smallest 
true hosts; this is because the hostless SNe lie in faint hosts that tend to also be small, either due to their intrinsic size 
and low-luminosity or because they are distant and thus subtend small angles.

Given the decreasing purity as a function of DLR separation seen in Figures~\ref{fig:MICE_EffPur} and \ref{fig:ACSGC_EffPur}, 
it is reasonable to ask if there is a value of DLR separation that we can use as a cut to remove probable mismatches. 
SNLS decided that SNe whose nearest galaxy is $>5R$ away do not 
get assigned a host, and we make a similar requirement using DLR.
To maintain an efficiency (true positive rate) of 98\%, we find that a cut at a distance of 5.3 DLR results in a purity of 94.45\% 
for MICECAT and removes 6.5\% of the sample. 
Similarly fixing the efficiency at 98\% for ACS-GC, we find that a cut at 11.5 DLR results in a purity of 97.29\% and removes 
7.1\% of the sample. These purity values are listed in Table~\ref{tab:matchres} for comparison.

\subsection{Comparison with Spectroscopically-Confirmed SNe in DES}
\label{Sec:DESY1Y2}

Host galaxy identification in DES is performed using the DLR method within an initial 15\arcsec\ search radius around 
each transient\footnote{For our simulations, we find that a cut on SN-host separation of 15\arcsec\ removes 0.3\% of SNe 
in MICECAT and 1.4\% of SNe in ACS-GC.}. 
The DLR for nearby galaxies is currently computed from the SExtractor parameters {\tt A\_IMAGE}, 
{\tt B\_IMAGE}, and {\tt THETA\_IMAGE} obtained from the co-added $r+i+z$ detection images taken during 
Science Verification (``SVA1"). In the future, we plan to create deeper multi-season co-added images without SN light to use 
for host galaxy identification and host studies. 

To test the DLR method for DES-SN, we examine the sample of spectroscopically-confirmed SNe 
discovered in DES Years 1 and 2 and estimate the accuracy of the host matching based on the agreement between the redshift 
obtained from the SN spectrum and the redshift obtained from an independent spectrum of the galaxy we identify as the host. 
Of the 106 SNe (of all types) with spectral classifications, 73 also have a spectrum of the host galaxy. Two of those 73 have 
SN redshifts that disagree with the host redshifts by more than 0.1, indicating the host was misidentified. Of the remaining 71, 
the difference between the SN redshift and the host redshift is at most 0.021, with a mean and standard deviation of 0.0017 
and 0.0054, respectively. This indicates very good agreement and a high likelihood of a correct host match, though in cases 
of SNe in galaxy groups or clusters the redshift agreement between the SN and any cluster member will be similarly good. 
Furthermore, for 8 cases out of these 71 the host galaxy is not the nearest galaxy in angular separation, and all but one of 
those nearest galaxies lacks a spectroscopic redshift to compare to the SN redshift. However, for one case there exists 
galaxy redshifts for both the host (nearest galaxy in DLR-space) and the nearest galaxy in angular separation, 
and these redshifts differ by only 0.0002, which is evidence that these two galaxies belong to the same group or cluster. 
This single example illustrates the difficulty in host identification. For this reason, we advocate that for the cases where the 
nearest DLR galaxy is different from the nearest angular separation galaxy that both galaxies be targeted for spectroscopic 
follow-up. Having redshifts of both galaxies is necessary to better quantify the rate of occurrence of SNe in high-confusion 
regions such as galaxy groups and clusters. 

From this DES sample we can roughly estimate the host galaxy mismatch rate due to the failure of the DLR method to be 
2.7\% (2/73). We compare this rate 
to the $\sim 3-5\%$ DLR failure rate from our simulations (where we have ignored the hostless SNe). Of course, this sample 
of spectroscopically-confirmed SNe with host redshifts is highly biased, since both the SNe and hosts must be bright 
enough to be targeted and to obtain secure redshift measurements. A description of the first 3 years of the DES 
spectroscopy campaign to target live transients and their host galaxies will be published in D'Andrea et al., in prep.

\subsection{Implications for Cosmology}
\label{Sec:CosmoImp}

\begin{figure*}[!tb]
\epsscale{1.2}
\plotone{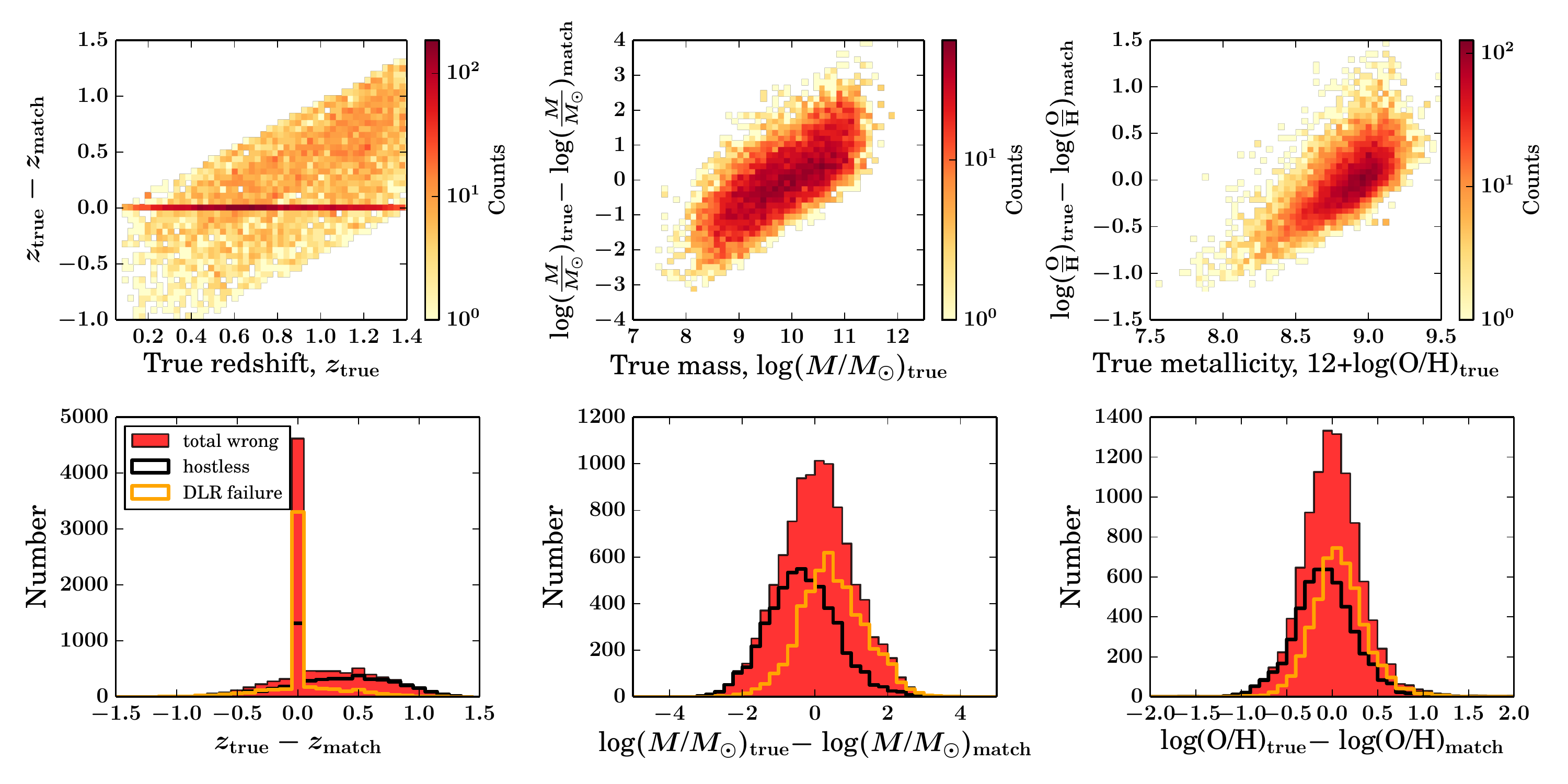}
\caption{The difference in galaxy properties between the true host and the matched host for the 
wrong matches (including hostless SNe) among the 100K SNe simulated on MICECAT galaxies. 
Plots show (from left to right) redshift, stellar mass, and metallicity.}
\label{fig:MICE_DeltazMmet}
\end{figure*}

Since the MICECATv2.0 galaxies all have redshifts, stellar masses, and gas-phase metallicities, 
we can investigate host galaxy mismatches as a function of these key host properties which influence 
cosmological inferences obtained from SNe. Figure~\ref{fig:MICE_DeltazMmet} displays the differences between 
the true and matched galaxy in terms of redshift, mass, and metallicity for cases where there is a host mismatch.  
The data plotted are for the $\approx 10,000$ wrong matches out of the 100K simulated SNe on MICECAT host galaxies. 

\begin{figure*}[!tb]
\epsscale{0.8}
\plotone{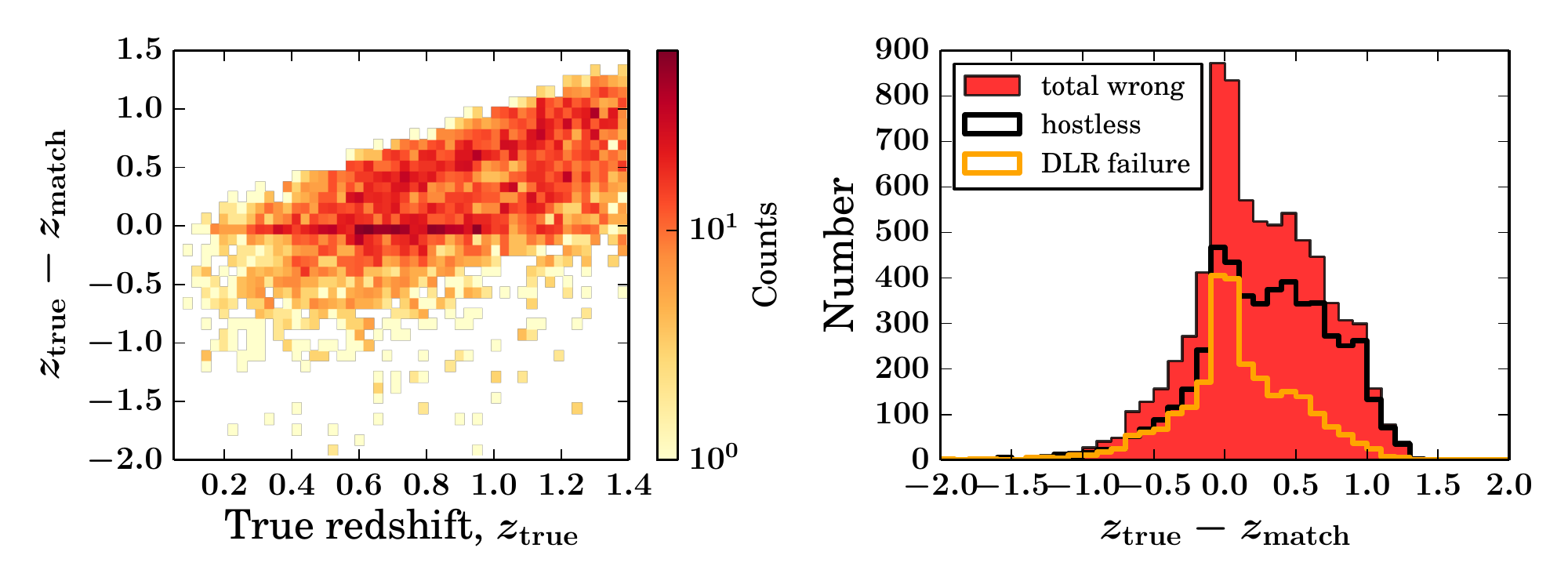}
\caption{The difference in redshift between the true host galaxy and the matched host galaxy for the  
wrong matches among the 100K SNe simulated on ACS-GC galaxies (for which both the true host and the 
matched host have redshifts listed in the catalog).}
\label{fig:ACSGC_Deltaz}
\end{figure*}

The distribution of redshift differences, $z_{\text{true}} - z_{\text{match}}$, is highly peaked at zero, indicating that the 
mismatched galaxy is often at a very similar redshift as the true host and is likely a group or cluster neighbor.  This is 
encouraging given the reliance on host redshifts for SN classification and placement on the Hubble diagram. 
However, the distribution of redshift differences has large tails which are asymmetric, indicating that for hostless SNe the  
mismatched galaxy is more likely to be a lower-redshift foreground galaxy. This makes sense given that the hostless
fraction rises with increasing redshift (upper right panel, Figure~\ref{fig:MICE_EffPur}). 
Given the known Hubble residual correlation with host-galaxy mass, current cosmological analyses with SNe Ia use the 
host mass to correct SN luminosities \citep[e.g.,][]{sul11,JLA}. 
Using the mass of the wrong galaxy may cause an incorrect offset to be applied to the SN peak magnitude. 
There is also some theoretical evidence that the true driver of this effect is SN progenitor metallicity \citep{tim:03,kas:09} 
or age \citep{chi14}.  
For these reasons we include both host stellar mass and gas-phase metallicity in Figure~\ref{fig:MICE_DeltazMmet}, 
as these parameters (but not galaxy age) are included in MICECATv2.0.

For all galaxy properties shown, the differences can be extreme ($\Delta z \sim 1$,  $\Delta (\log{M}) \sim 3\unt{dex}$, 
$\Delta (\log{[\mathrm{O/H}]}) \sim 1\unt{dex}$), which is disconcerting. The distributions of mass and metallicity 
differences, shown in the lower panels of Figure~\ref{fig:MICE_DeltazMmet}, are much broader than the redshift 
difference though the total wrong-match distributions still peak at zero. The location of this peak will shift depending 
on the ratio of hostless SNe to DLR failures. If we examine the breakdown of the total wrong-match 
histogram, we notice that the DLR failures are biased to be greater than zero while the hostless 
cases are biased to be less than zero. This is because the hostless SNe are generally low-mass and 
low-metallicity (as well as faint) and so are more likely to get mismatched to galaxies with higher masses and 
higher metallicities. Similarly, for the DLR failures (the brighter true hosts), the true hosts tend to be higher 
mass/metallicity, so the likelihood of the SN getting mismatched to galaxies of lower mass/metallicity is higher.  

As previously mentioned, several recent cosmological analyses have used a ``mass step" correction to SN luminosities 
such that SNe Ia in hosts with $\log(M/M_\odot) \leq 10$ have one absolute magnitude and those in hosts with 
$\log(M/M_\odot) > 10$ have another \citep{sul11,JLA}. 
Using the MICECAT sample of mismatched SNe, we can ask how often a SN gets matched to a host galaxy that 
falls into a mass bin that is different from the mass bin of the true host. That is, how often is it that a SN in a truly low-mass 
host gets matched to a high-mass galaxy, or that a SN in a truly high-mass host gets matched to a low-mass 
galaxy? Using a split value of $\log(M/M_\odot) = 10$, as done in the literature, to separate low- and high-mass galaxies 
(the MICECAT true host galaxy mass distribution has a median of 10.163), we find that this occurs 44\% of the time. 
Given that the total mismatch rate is $\approx 10\%$, this implies that $>4\%$ of the total SN Ia sample would be 
assigned an incorrect luminosity and thus be misplaced on the Hubble diagram. 

The ACS-GC catalog does not contain galaxy mass or metallicity estimates but does contain spectroscopic or photometric 
redshifts for the majority of galaxies.  Therefore, of the 100K simulated SNe on ACS-GC host galaxies, we make a 
plot similar to Figure~\ref{fig:MICE_DeltazMmet} for the $\approx 7500$ incorrectly-matched SNe that have 
redshifts for both the true host and the matched host.  This is shown in Figure~\ref{fig:ACSGC_Deltaz}. 
While the redshift difference distribution is still peaked at zero as it is for MICE, the peak is not nearly as sharp.  
This plot also exhibits an asymmetry, indicating that SNe are more often mismatched to galaxies with redshifts lower 
than the true redshift. The exact shape of this redshift difference distribution depends on the redshift distribution 
of detected SNe and the magnitude limit of the survey, among other factors. 

Since there is clearly a redshift dependence of the matching accuracy, we emphasize that this could be 
potentially problematic since relative distances of SNe Ia are used to infer cosmological parameters.  
Although a detailed analysis is beyond the scope of this paper, the possibility of misclassified SNe as 
well as mismatched host galaxies must be accounted for in cosmology frameworks \citep[e.g.,][]{rub15}.

\section{IMPROVEMENTS USING MACHINE LEARNING}
\label{Sec:ML}

While the automated DLR algorithm presented in Section~\ref{Sec:HostMatching} is $90-92\%$ accurate at 
matching SNe to their proper host galaxies, for real data we will not know the identity of the true host. 
The algorithm produces a match but does not produce an uncertainty or a probability that an individual SN-host matched 
pair is correct. Therefore, we would like some way of quantifying the likelihood of a correct match for each SN, while at the 
same time improving the matching accuracy. 

In order to do this, we employ machine learning (ML) to compute probabilities that can be used to classify 
our SN-host matched pairs into two classes -- ``correct match" and ``wrong match."
Our goal is to create a binary ML classifier that uses features of the data extracted from the 
results of the matching algorithm to quantify the probability of a correct host match for every SN.
We use a Random Forest \citep[RF;][]{RF} classifier since this method is fast, easy to implement, and was 
successfully used by \citet{autoscan} to train a binary classifier to separate artifacts from true transients in DES SN 
differenced images. RF is also capable of providing probabilities for class membership, which in effect tells us the 
likelihood that a SN-host matched pair is correctly matched (i.e. belongs to class ``correct match").\footnote{However, 
we note that the RF probabilities must first be calibrated before being used in a likelihood analysis.} 
We use the RF implementation available in the Python package {\tt scikit-learn} \citep{scikit-learn}.

We describe the features we use in Section~\ref{Sec:Features} and introduce our binary ML classifier in 
Section~\ref{Sec:RFClassify}. In Section~\ref{Sec:Optimize} we explain how we train and optimize the classifier, 
and finally we present the results in Section~\ref{Sec:MLResults}.

\subsection{Features: Distinguishing Correct \& Wrong Matches}
\label{Sec:Features}

As described in Section~\ref{Sec:HostMatching}, host galaxy matching begins by considering galaxies within a search 
radius around the SN position. As part of the matching algorithm, distances from the SN position to each 
potential host are measured in units of DLR (\ddlr).  
Let us adopt the shorthand notation for the \ddlr\ of the $i$th host as $D_i$ and then order the potential hosts by 
increasing DLR such that $D_1$ is the value of \ddlr\ for the nearest galaxy in DLR-space.
Similarly, let us denote $S_i$ as the angular separation (in arcsec) of the $i$th host from the SN such that when 
ordered by increasing angular separation, $S_1$ is the nearest galaxy in angular-space.  

Confusion over the identification of a host galaxy will occur in situations where nearby galaxies have similar 
separations from the SN, creating ambiguity over which is the true host.  Therefore, we would expect that 
$D_i$ and functions thereof, such as $D_i - D_j$ or $D_j/D_i$, have different distributions for correct and wrong 
matches; the same ought to be true for $S_i$ and functions thereof.  
In most cases, this host ambiguity exists between the nearest galaxy (with separation 
$D_1$) and the second-nearest galaxy (with separation $D_2$).  
As a result, values of $D_2 - D_1$ or $D_1/D_2$ are good indicators of whether or not a SN was correctly 
matched to a host galaxy.  We refer to such indicators as features of the host-matched data.

A more revealing feature is the difference in \emph{angular} separation between the SN and the nearest DLR galaxy, 
$S(D_1)$, and the SN and the second-nearest DLR galaxy, $S(D_2)$.  
Let us call this $\Delta S(D21)$ and define it as $\Delta S(D21) = S(D_2) - S(D_1)$. This feature has the interesting 
property of being a combination of DLR and angular separation.  In most cases, matching using the DLR method as 
we have done will select the same host galaxy as matching by simply taking the nearest galaxy in angular separation. 
For these cases, the host is the galaxy with minimum \ddlr\ ($=D_1$) and minimum angular separation ($S_1$), and so 
$\Delta S(D21) > 0$.  However, for cases where the DLR method and the angular separation method disagree, negative 
values of $\Delta S(D21)$ are possible since the galaxy with minimum \ddlr\ ($D_1$) might actually be the 
second-closest galaxy in angular separation ($S_2$).  Therefore, cases where $\Delta S(D21) < 0$ have a higher  
chance of being incorrect matches.  

\begin{figure}[tb]
\epsscale{1.2}
\plotone{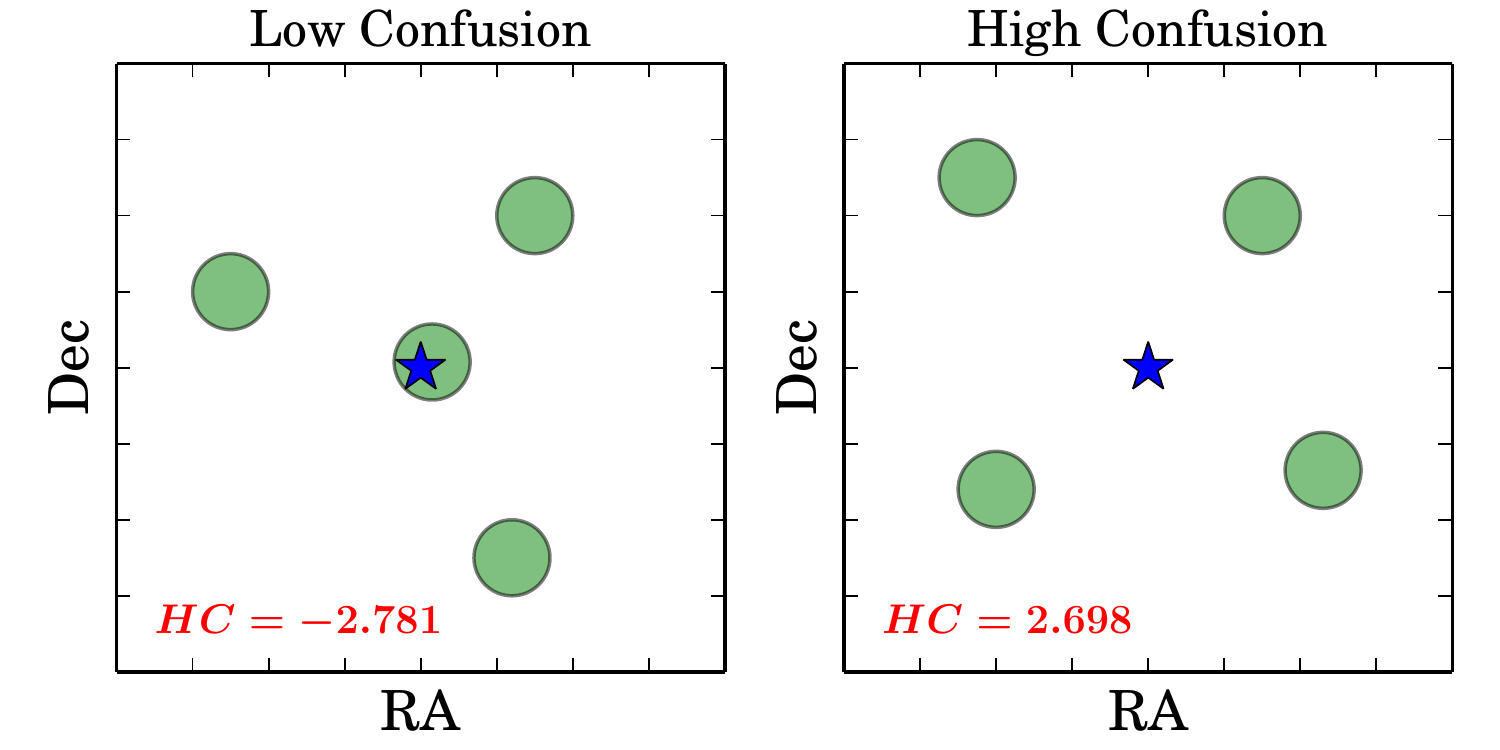}
\caption{An illustration of the difference between cases of low host confusion (left) and high host confusion (right).  
In both cases, the star in the center represents the position of the SN, 
and the circles represent nearby galaxies, projected on the sky.  
For simplicity of this example, all galaxies are depicted as circles of the same size and thus all have the same DLR. 
However, as their angular distances from the SN differ, they will have different values of \ddlr. 
The respective values of the host confusion parameter, $HC$ (see Equation~\ref{eqn:HC}), are shown on each panel.}
\label{fig:HCex}
\end{figure}

We aim to define a quantity that parametrizes the degree of host confusion or 
mismatching for a given SN in such a way that a larger value indicates a higher degree of confusion.  
Given a SN location and $N$ galaxies within our search radius, we define a host confusion parameter, 
$HC$, to be 
\begin{equation}\small
HC =
  \begin{dcases}
   -99 & \text{if } N = 1 \\
    \log_{10}\lr{\frac{D_1^2/D_2 {+} \epsilon}{D_2 {-} D_1 {+} \epsilon} 
    \sum_{i=1}^{N-1}\sum_{j>i}^{N}{\frac{D_i/D_j {+} \epsilon}{i^2(D_j {-} D_i {+} \epsilon)}}}
    & \text{if } N > 1. 
  \end{dcases}
\label{eqn:HC} 
\end{equation}
The sum is over all pairs of galaxies within the search radius and accounts for cases where any number of the $N$ 
nearby galaxies have similar separations from the SN. The prefactor term outside the sum increases the contribution from 
the two nearest galaxies, which generally cause the majority of the confusion.  
The $D_1/D_2$ term in the numerator reduces the overall value of $HC$ for cases where $D_1$ is small but $D_2$ is large 
by comparison; the extra factor of $D_1$ in the numerator penalizes SNe which are far separated from their hosts. 
The $D_2-D_1$ term in the denominator increases the value of $HC$ for cases where the first and second 
DLR-ranked galaxies are very close in separation ($D_1 \approx D_2$). The addition of a small quantity, $\epsilon$, 
prevents $HC$ from becoming undefined or infinite in cases where $D_i=0$ or $D_i=D_j$. 
We choose $\epsilon = 10^{-5}$, but find the values of $HC$ to be relatively insensitive to the precise value of $\epsilon$.
Inside the sum, the $i^2$ term is a weight factor 
that progressively down-weights the contributions from galaxies as they get farther away from the SN, 
the rationale being that the more distant galaxies are less likely to contribute to the confusion. 
$HC$ has the desired general behavior of being small when the differences between the potential hosts are large 
(low density, low degree of confusion) and large when these differences are small (high density, high 
degree of confusion).  A cartoon illustrating the difference between cases of low and high confusion is shown in 
Figure~\ref{fig:HCex}. 

\begin{figure*}[!tb]
\epsscale{0.8}
\plotone{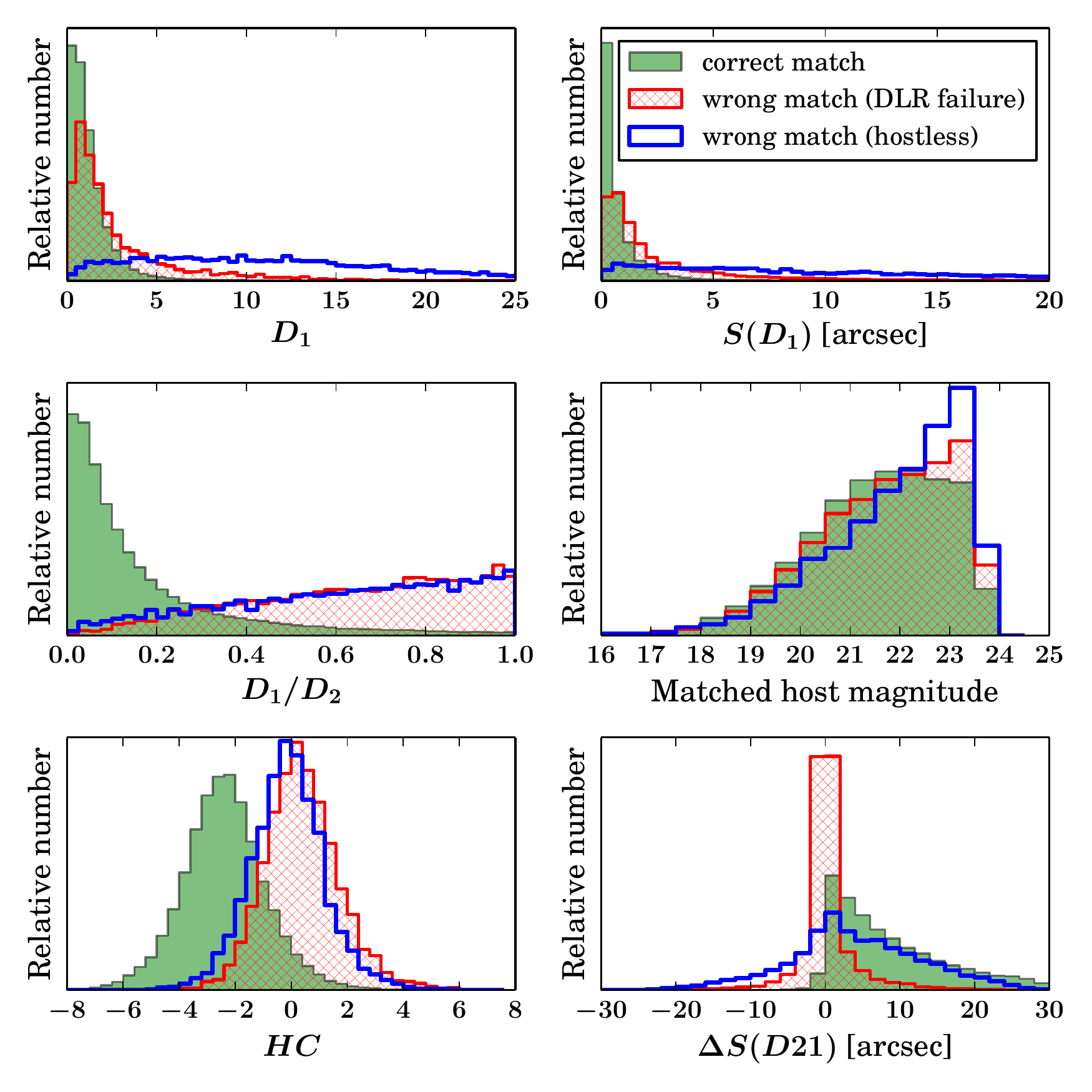}
\caption{Distributions of a subset of the features derived from the results of our host-matching algorithm run on SNe 
simulated on MICECAT galaxies.  These features show the difference in distributions between 
correct matches (green filled), wrong matches due to failures of the DLR matching algorithm (red cross-hatched), 
and wrong matches due to the SNe being hostless (blue). The area of each histogram is normalized to unity.}
\label{fig:FeaturesMICE}
\end{figure*}

\begin{figure*}[!tb]
\epsscale{0.8}
\plotone{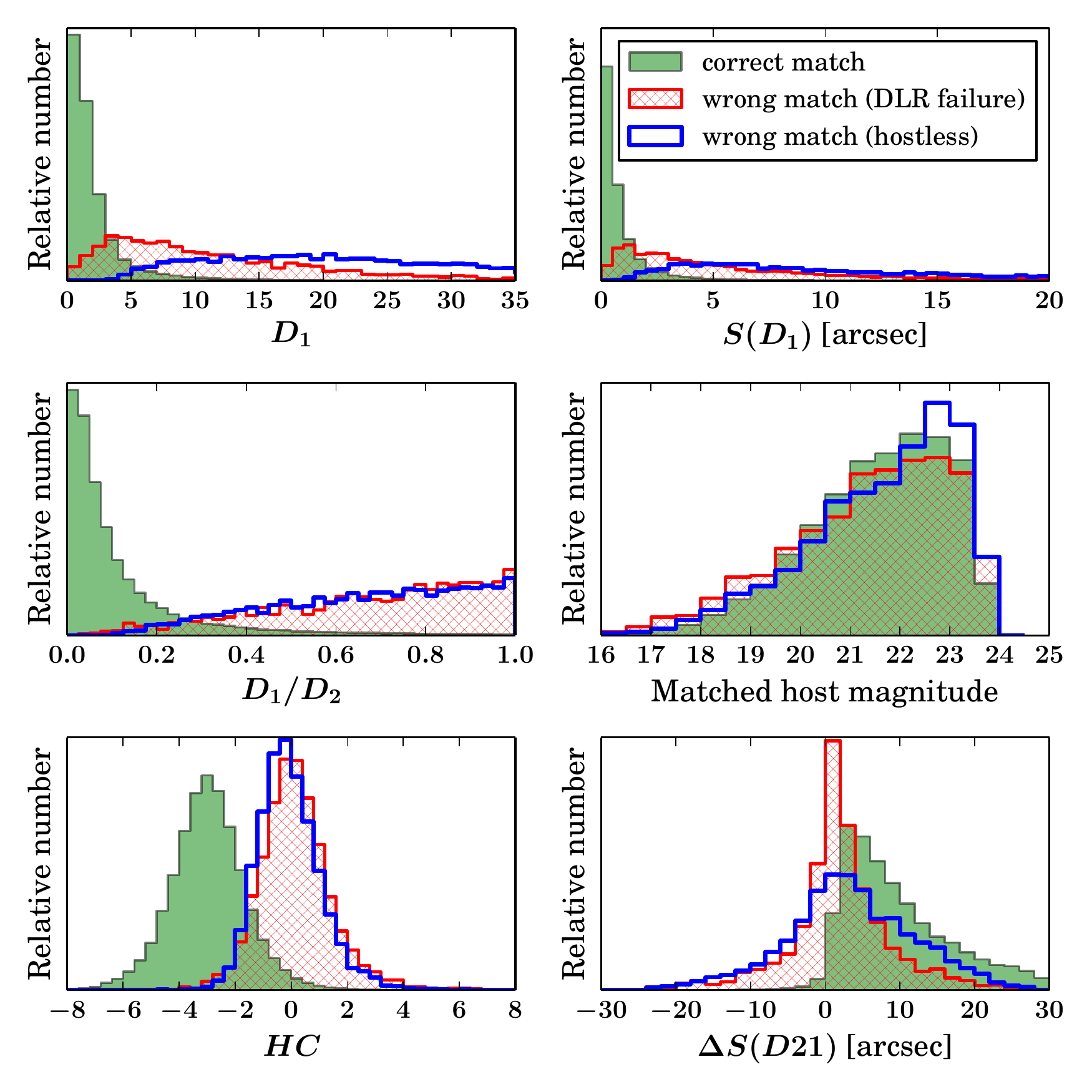}
\caption{Same as Figure~\ref{fig:FeaturesMICE} but for the SNe simulated on and matched to the ACS-GC galaxy catalog.}
\label{fig:FeaturesACS}
\end{figure*}

The distributions of $HC$ for both correct and wrong host-galaxy matches as well as hostless SNe are plotted in 
Figure~\ref{fig:FeaturesMICE} (MICE) and Figure~\ref{fig:FeaturesACS} (ACS-GC) along with a subset of the other features 
that we have described above.  Ideally, we would like to see clear separations in the distributions of features
between correct matches (shown in green filled) and the incorrect matches, which include matches that are wrong due to
a failure of the DLR method of matching (shown in red cross-hatched) and also hostless cases (shown in blue).  
The hostless matches 
will be wrong by construction since these SNe were simulated on faint galaxies that are then removed by the magnitude 
limit during the matching process. However, we would hope that the hostless distributions are more similar to the wrong 
match distributions than to the correct match distributions. Given an actual observed SN, we would like to be aware 
if there is a high probability that its matched host is wrong, whether due to host confusion or due to the true host being 
low-luminosity (hostless). 

Indeed, the hostless distributions for the features shown in Figures~\ref{fig:FeaturesMICE} and 
\ref{fig:FeaturesACS} differ significantly from the correct match distributions. In addition, the hostless and DLR failure 
distributions are very similar in general, which is promising. 
The distribution of $D_1/D_2$ is very similar for MICECAT and ACS-GC, as is the distribution of $\Delta S(D21)$, 
although the latter distribution is broader for ACS-GC. 
An interesting difference between MICECAT and ACS-GC is seen in the $D_1$ and $S(D_1)$ feature distributions. 
For MICECAT, the DLR failures for these features look much like correct matches, while for ACS-GC the DLR failures 
are well-separated from correct matches. This might be a clue toward explaining the overall higher matching 
accuracy in ACS-GC compared to MICECAT, the origin of which is explored in the Appendix. 

Additional features of the data can always be discovered or developed and included into the ML training to 
improve performance. Other potentially useful features worth exploring in the future include SN photo-z, 
photometrically-determined SN type, and host galaxy morphological type.
Furthermore, surveys like DES also have photo-z estimates of all galaxies in the survey area. 
These in conjunction with SN photo-z could be used in the matching process, either as weighted probability 
densities or as input features for ML.

\subsection{Binary Classification with Random Forest}
\label{Sec:RFClassify}

\begin{figure}[tb]
\epsscale{0.9}
\plotone{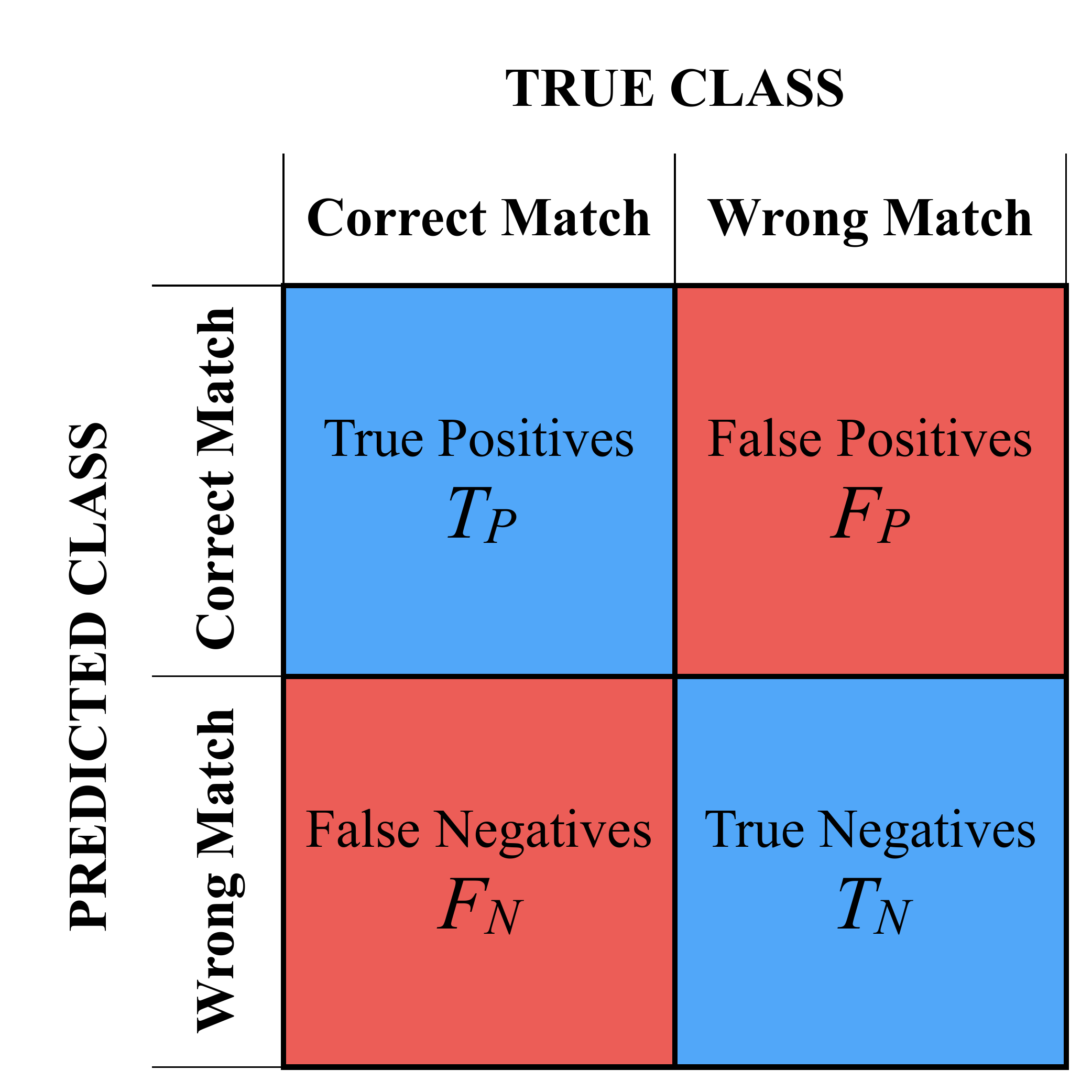}
\caption
{A diagram of the confusion matrix for binary classification into classes ``correct match" and ``wrong match".}
\label{fig:ConfMat}
\end{figure}

For the task of binary classification, as we have here, it is useful to consult the schematic $2 \times 2$ confusion matrix 
shown in Figure~\ref{fig:ConfMat}.  Objects that are correct matches (i.e., belong to the true class ``correct match") 
and which the classifier predicts are correct matches are called true positives ($T_P$); those that are correct matches but 
are predicted to be wrong matches are called false negatives ($F_N$). Objects that are wrong matches (true class 
``wrong match") are called false positives ($F_P$) if they are predicted to be correct matches and are called true negatives 
($T_N$) if they are predicted to be wrong matches. 

Using these definitions, we can also define the efficiency and purity of the classifier. 
Efficiency is given by   
\begin{equation}
\text{efficiency} = \frac{T_P}{T_P + F_N}
\end{equation}
and is also known as the true positive rate.  The efficiency is the fraction of true correct matches recovered by the classifier. 
Purity is defined as
\begin{equation}
\text{purity} = \frac{T_P}{T_P + F_P}
\end{equation}
and is essentially the accuracy with which objects are classified as correct matches.
The results of the host-matching algorithm can be thought of as having an efficiency of 100\% (since all SNe get 
matched to a host galaxy) but with a purity of $<100\%$ (since some fraction of those matches will be incorrect). 
The goal of this ML classifier is to increase the purity (matching accuracy) of the SN-host galaxy matched 
sample, with some minimal decrease in efficiency. In this way, we lose some SNe but become more confident in the 
accuracy of the host galaxy matches for those SNe that remain. 
For a more comprehensive description of machine learning with RF, see \citet{RF} and \citet{autoscan}. 

RF can output probabilities of a correct match, $P_{corr}$, for each SN-host pair in the test sample.
Classification into ``correct match" or ``wrong match" depends on the threshold probability, $P_t$, which is the probability 
above which a SN-host pair is classified as a correct match. The value of $P_t$ can be selected to maximize the metric 
of choice, such as efficiency or the purity, and depends on the scientific goals. 
For example, if a SN survey requires that no more than 2\% of correct matches be misclassified (i.e., false negative 
rate $= 2$\%), then one would choose the value of $P_t$ at which the efficiency (= $1-$ false negative rate) 
equals 98\% and compute the corresponding purity. For this study, we select as our metric the value of purity at a 
fixed efficiency of 98\%.

\subsection{Training and Optimization}
\label{Sec:Optimize}

Our RF classifier must first be trained in order to learn how to properly classify SN-host pairs into ``correct match" 
and ``wrong match" classes. While the majority of matches determined from our DLR matching algorithm are correct 
(see Section~\ref{Sec:MatchResults}), we also have cases of mismatched pairs due to failures of the DLR method and 
hostless SNe.  A training sample containing a realistic proportion of correct and wrong matches (roughly 10:1) would 
bias the classifier, since it would not have enough examples of wrong matches to learn how to distinguish them from correct 
matches.  Therefore, to reduce this bias we attempt to evenly balance the training set so that it contains equal 
numbers of correct and wrong matches. The training set of ``wrong matches" comprises both misidentification 
due to failure of the DLR method and misidentification of hostless SNe, in the proportion they appear in the data 
(given the 5\% hostless rate assumed in Section~\ref{Sec:MagLim}). Training is performed separately for MICECAT and 
ACS-GC datasets. Each classifier is trained on equal numbers of correct and wrong matches taken from the 100K 
simulated SNe from Section~\ref{Sec:HostMatching}. The training sample size for MICE is $\approx 20$K while for 
ACS-GC it is $\approx 15$K. 

A Random Forest is constructed from a user-defined set of parameters called hyperparameters that control the growth 
and behavior of trees in the forest.  The Random Forest 
implementation we use relies on the following hyperparameters: 
\begin{enumerate}
\item {\tt n\_estimators}, the number of decision trees in the forest
\item {\tt criterion}, the function used to measure the quality of a split at each node 
\item {\tt max\_features}, the maximum number of features considered when looking for the best split at a node
\item {\tt max\_depth}, the maximum depth of a tree 
\item {\tt min\_samples\_split}, the minimum number of samples required to split an internal node.
\end{enumerate}

We optimize our RF classifier by varying these hyperparameters over the range of values listed in 
Table~\ref{tab:hyperpar}.  We performed a 3-fold cross-validated randomized search, sampling 1000 random points 
over this hyperparameter space.  
For {\tt n\_estimators}, {\tt max\_features}, and {\tt min\_samples\_split} we randomly select integer values from the 
uniform distributions given by ($min, max$) in Table~\ref{tab:hyperpar}.  For {\tt criterion} and {\tt max\_depth} we 
randomly sample from the discrete possibilities listed in brackets. 
The performance metric of the classifier was defined to be the value of purity at an efficiency of 98\%.  
Combinations of hyperparameters that maximize this metric were 
considered optimal for our purposes. The performance metric can be chosen by each SN survey 
to meet the needs and goals of the survey and need not be the same as the one we chose here.

We find that the entropy criterion consistently outperformed the Gini criterion\footnote{Entropy uses information gain 
as the metric while Gini uses the Gini impurity.}, and that performance is  
insensitive to the values of {\tt max\_depth} and {\tt min\_samples\_split}. Performance increases for values of 
{\tt n\_estimators} up to $\sim 100$ and then plateaus for larger values. Similarly, performance increases for values of 
{\tt max\_features} up to 4 and then plateaus for larger values. 
Therefore, we select the following as our hyperparameters when implementing our RF for classification:  
{\tt n\_estimators}=100, {\tt criterion}=entropy, {\tt max\_features}=10, {\tt max\_depth}=None, 
and {\tt min\_samples\_split}=70. These values are also listed in Table~\ref{tab:hyperpar}.

\begin{deluxetable}{lcr}
\tablecaption{Random Forest Hyperparameter Values for Optimization \label{tab:hyperpar}}
\tablewidth{0pt}
\tablehead{
\colhead{Hyperparameter} &
\colhead{Range} & 
\colhead{Selected} 
}
\startdata 
{\tt n\_estimators} & $(10, 300)$ & 100 \\
{\tt criterion} & \{gini, entropy\} & entropy \\
{\tt max\_features} & (1, 11) & 10 \\
{\tt max\_depth} & \{None, 15, 30, 50, 80\} & None \\
{\tt min\_samples\_split} & (10, 100) & 70 \\
\enddata
\tablecomments{For ranges denoted in parentheses, integer values were randomly sampled from the uniform 
distribution ($min, max$). For ranges denoted in braces, random values were selected from the discrete options listed. 
The values eventually used in the Random Forest classifier are listed under the column ``Selected."}
\end{deluxetable}

\subsection{Results \& Performance}
\label{Sec:MLResults}

Here we present the results from our ML classifier on SN-host matched pairs. After training, the relative importance of 
the features used in the training sample can be computed. The general method used to compute RF feature importances 
is described in Section 3.4 of \citet{autoscan}. 
The importance of a feature is a number such that a higher value indicates the feature is more relevant in providing 
information during training.  The importances are normalized so that they sum to unity. 
In Table~\ref{tab:importances}, we list all the features used to train our classifiers and give their relative importances 
for both MICE and ACS-GC.  By far the most important feature for both MICECAT and ACS-GC is $D_1/D_2$,  
with importances $> 0.5$. The second most important feature in both cases is $D_1$. 
For ACS-GC, all other feature are nearly irrelevant (with importances $< 0.04$), whereas for MICE the other features 
help contribute more toward the classification. The feature $\Delta S(D21)$ is important for MICECAT but not so 
for ACS-GC. Our derived feature, $HC$, is the fourth most important feature in the ML training process for both MICECAT 
and ACS-GC.

\begin{deluxetable*}{lcccc}
\tablecaption{List of Machine Learning Features \label{tab:importances}}
\tablewidth{0pt}
\tablehead{
\colhead{Feature} & \multicolumn{2}{c}{MICECATv2.0} & \multicolumn{2}{c}{ACS-GC COSMOS}\\
\colhead{} & \colhead{Importance} & \colhead{Rank} & \colhead{Importance} & \colhead{Rank}
}
\startdata 
$D_1$ & 0.114 & 2 & 0.179 & 2\\
$S(D_1)$ & 0.056 & 5 & 0.016 & 5 \\
$\Delta S(D21)$ & 0.083  & 3 & 0.011 & 8 \\
$D_2 - D_1$ & 0.024 & 8 & 0.011 & 9 \\
$D_1/D_2$ & 0.525 & 1 & 0.685 & 1 \\ 
$D_3 - D_1$ & 0.010 & 11 & 0.008 & 11 \\
$D_1/D_3$ & 0.012 & 10 & 0.033 & 3 \\
$HC$ & 0.065 & 4 & 0.017 & 4 \\
MAG (matched galaxy magnitude) & 0.053 & 6 & 0.013 & 7 \\
$A$ (matched galaxy size) & 0.039 & 7 & 0.010 & 10 \\
$B/A$ (matched galaxy axis ratio) & 0.018 & 9 & 0.015 & 6 \\
\enddata
\tablecomments{Feature importances and ranks computed from a single training. Importances will fluctuate slightly after 
each random training.}
\end{deluxetable*}

\begin{figure*}[tb]
\epsscale{1}
\plotone{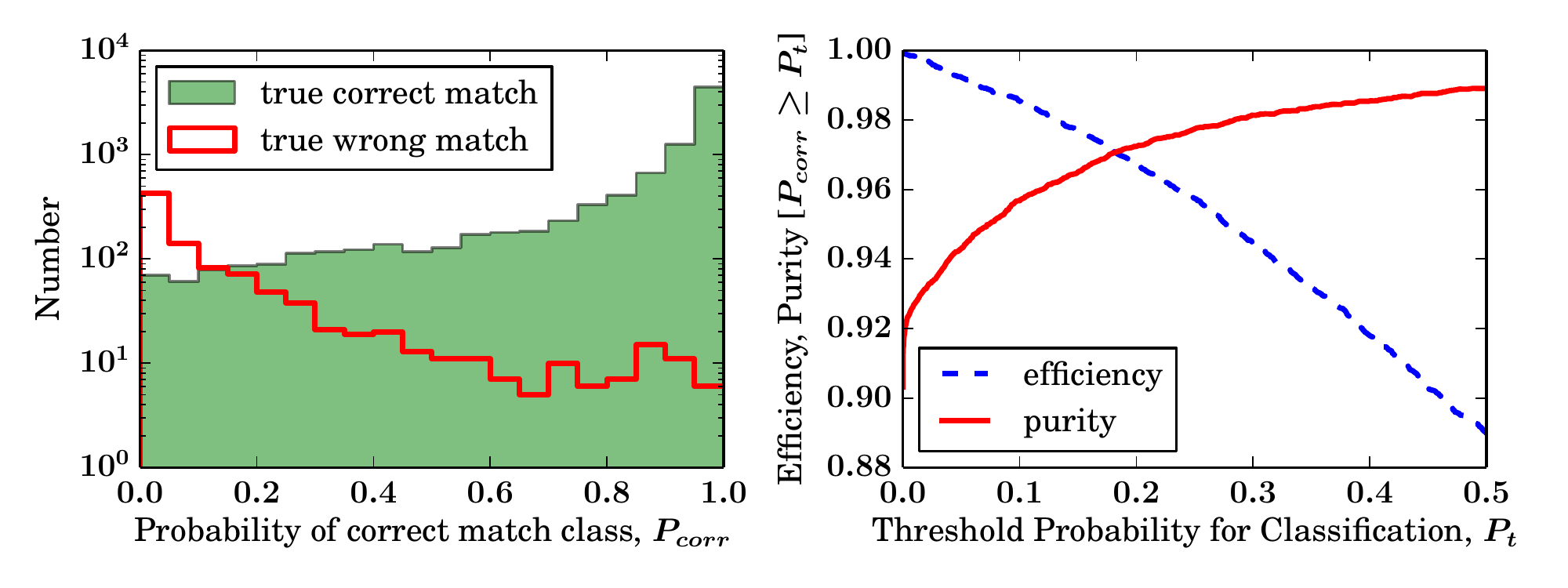}
\caption
{Results of the ML classifier on a validation set of 10,000 SNe simulated on galaxies from MICECATv2.0. 
\emph{Left}: The ML probability of a SN-host pair being a correct match, with the true correct matches 
shown as the green filled histogram and the true wrong matches (including ``hostless" SNe) shown as the red 
open histogram. Note the logarithmic scaling of the ordinate axis.
\emph{Right}: The efficiency and purity as a function of ML threshold probability. SN-host pairs with 
probabilities $P_{corr}>P_t$ get classified as correct matches.}
\label{fig:MICE_EffPurML}
\end{figure*}

\begin{figure*}[tb]
\epsscale{1}
\plotone{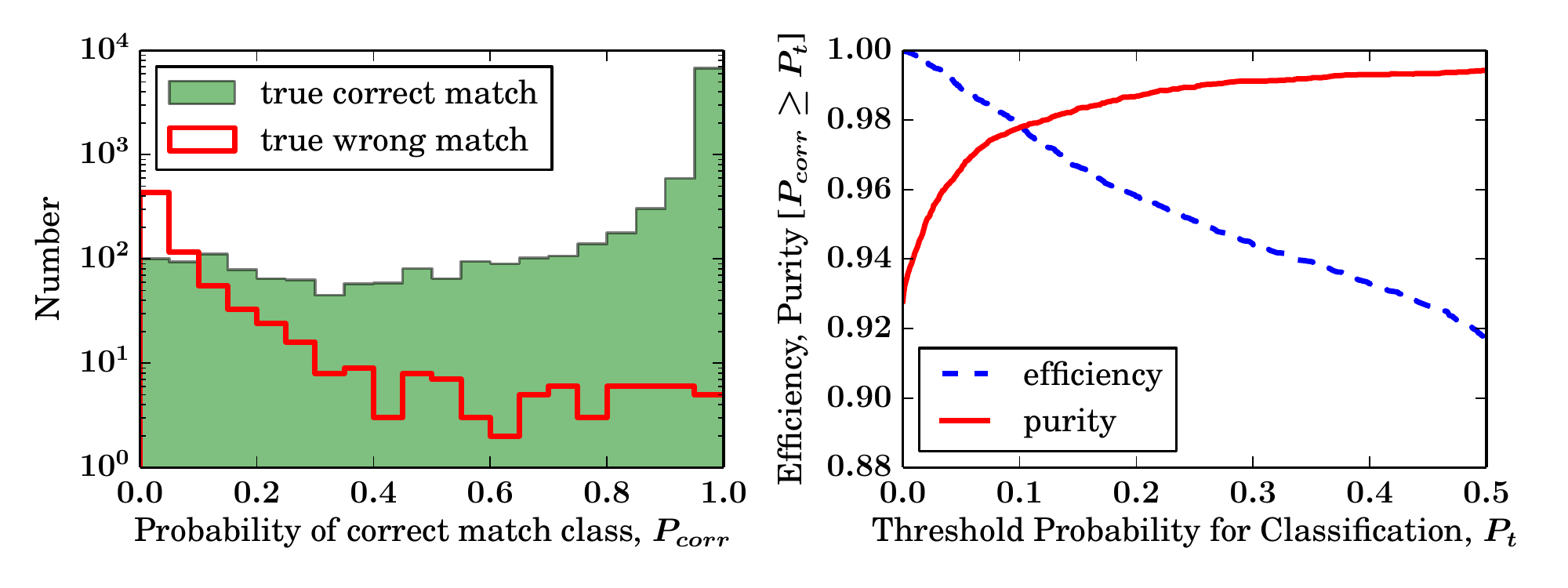}
\caption
{Same as Figure~\ref{fig:MICE_EffPurML}, but for SNe simulated on ACS-GC galaxies.}
\label{fig:ACSGC_EffPurML}
\end{figure*}

To demonstrate the improvement that ML provides here, we apply our classifier to an independent validation set of 
simulated SNe (10K each for MICECAT and ACS-GC) that were matched to hosts via the DLR method, again with 
5\% of these SNe being hostless. Figures~\ref{fig:MICE_EffPurML} and \ref{fig:ACSGC_EffPurML} show the results 
from MICECAT and ACS-GC, respectively. As before, the accuracy of the DLR matching algorithm before ML is 
90\% for MICE and 92\% for ACS-GC for the validation set, 
the same as the result seen with our 100K SNe (Table~\ref{tab:matchres}, first row). 

The left panels of Figures~\ref{fig:MICE_EffPurML} and \ref{fig:ACSGC_EffPurML} plot the ML output probability of being a 
correct match ($P_{corr}$), with the true correct matches shown in the green filled histogram and the true wrong matches 
(including hostless SNe) shown in the red open histogram. The ordinate axis displays number on a logarithmic scale. 
There is clearly a good separation between the two classes, with true wrong matches having probabilities near zero 
and true correct matches having probabilities near one, as desired. The right panels display the efficiency and purity of 
the classifier as a function of the threshold probability, $P_t$, which defines the boundary between the classes 
``correct match" and ``wrong match."  Under our requirement of fixed 98\% efficiency, we find that this results in a 
purity of 96.2\% for MICE and 97.7\% for ACS-GC. In the right panels in both figures we see the dramatic increase of purity 
(matching accuracy) resulting from ML run after the initial matching algorithm. 
A summary of these results is provided in the last row of Table~\ref{tab:matchres}. We see that ML improves the purity 
above that of a simple cut on DLR separation, especially in the case of MICECAT. 
Similar to the cut on separation, this increase in purity with ML results in $7-8\%$ 
of the total SN sample being classified as having wrong matches. If a SN survey decides to remove these wrong matches in 
an analysis, it would constitute a significant reduction in sample size.  

However, a cut on DLR separation can only accept or reject a host match whereas ML is able to provide probabilities of a 
correct match. We wish to point out that the end result need not be binary classification into ``correct match" or ``wrong match." 
In the work we have presented, the binary classification was made based on the selection of a threshold probability 
that provides 98\% efficiency. SN-host matches that fall below this threshold are classified as ``wrong matches" and 
those above are classified as ``correct matches."
However, as the actual ML classifier outputs are the probabilities themselves, one could instead use the (calibrated) 
probabilities as weights in a Bayesian cosmology analysis and avoid binary classification and the outright rejection of SNe from 
the sample due to host misidentification. 

The ML classifier is specific to the dataset being used and so feature distributions and 
importances will vary between datasets (this is evident from comparing Figures~\ref{fig:FeaturesMICE} and 
\ref{fig:FeaturesACS}). Therefore, before we can apply this ML classifier to real SN data from DES, for example, 
it is critical that we first train the classifier on simulated SNe placed on galaxies in catalogs derived from real DES data. 
We leave such a DES-specific study for future work, since at this time we do not have adequate morphological 
classifications and light profile fits for DES galaxies. 
Furthermore, we have checked that using the nearest separation instead of the DLR as the initial host matching method, 
followed by an implementation of the ML classifier trained on analogous features (e.g., $S_1$, $S_1/S_2$, etc.) 
results in similar increases in purity.

\section{CONCLUSIONS}
\label{Sec:Conclusions}

In this paper we have investigated the problem of host galaxy identification, a challenge for modern SN surveys that must 
rely on host galaxies for SN cosmology.  For the DES SN Program this is a current concern, and the issue will be even 
more pressing for the LSST, which expects to discover hundreds of thousands of SNe Ia. 
Given limited resources to spectroscopically target all these SNe, host galaxy spectra will be the primary redshift source.  
We expand on the host matching algorithms published in previous works by testing our algorithm's efficacy with 
simulated SNe (including hostless SNe) and improving it with a machine learning classifier.

We have developed an automated algorithm that can be run on source catalogs and which matches SNe to 
host galaxies.  We have tested this algorithm by simulating SN locations on host galaxies in catalogs, both 
mock and real, and performing the matching using information on galaxies nearby the SNe.  
Using the DLR method of matching as outlined in Section~\ref{Sec:HostMatching} and assuming a hostless SN rate of 5\% 
results in a matching accuracy of $90-92\%$. Based on our simulations we find that the 
DLR method and the nearest angular separation method of matching select the same galaxy in the majority of cases. 
However, in the cases where these methods disagree, the DLR method is more often correct. 
This results in a statistically higher overall matching accuracy for the DLR method than simply matching hosts based on 
nearest angular separation. 

We have shown that the accuracy of host identification can be significantly improved with the addition of machine learning, 
which can be trained to output probabilities of a correct match. These probabilities, in turn, can be used to classify SN-host 
pairs into categories ``correct match" and ``wrong match," with purities as high as 97\% given a fixed 98\% efficiency. 
We find that regardless of the initial matching algorithm (DLR or angular separation), 
machine learning classification run afterward 
using features of the matched pairs does an excellent job of identifying probable correct and wrong matches.
We have also shown that the misidentification of host galaxies can result in values of redshift, mass, and metallicity 
that are very different from those of the true host. This in turn can result in the misplacement of SNe on the Hubble diagram. 

This work is intended as a proof of concept, illustrating an approach to host galaxy identification that can be applied to any 
SN survey. In order to apply this methodology to a given survey, several things are required. A large catalog of galaxies 
(preferably from real survey data) in the appropriate survey filters that contains positions, shapes, sizes, orientations,  
magnitudes, and light profiles is needed to place fake SN locations. In addition, having spectroscopic redshifts 
(or high-quality photometric redshifts) for as many galaxies as possible is useful if one wishes to simulate SNe with the 
same redshift distribution as the SN survey. A catalog generated from deep co-added images, corrected for 
seeing and not containing SN light will help reveal fainter galaxies and produce accurate shape measurements. 
SN locations simulated on these galaxies can then be matched using the same catalog and the match results used for 
training and validation sets for the machine learning classifier. 

The results presented come with several important caveats that we mention here. One is that we use 
a simple luminosity weighting rather than actual LFs for SN host galaxies from the literature, and so host galaxies that 
we select will not be completely representative of observed host galaxies of all SN types. 
Using SN data to better determine the distributions of SN-host galaxy separation for different types of SN, as opposed to 
using galaxy \sersic\ profiles to place SNe will improve studies of this kind. 
In addition, we do not account for observational  or instrumental factors such as SN detection efficiency and the PSF. 
For example, DES images in the SN fields have PSF sizes that are $>1\arcsec$, significantly larger than those of \hst\ 
and ACS-GC, which will make deblending and measurements of intrinsic galaxy sizes and shapes more challenging. 
Also, we assume a reasonable hostless SN rate of 5\% but the exact value will differ depending on the SN 
survey.

Future work is needed to implement the framework proposed here to determine 
the effect of host galaxy misidentification on cosmological parameters for a DES SN Ia analysis. 
This can be accomplished by simulating light curves of SNe Ia and core-collapse SNe onto galaxies actually observed 
in the DES SN fields and then running our host matching algorithm and machine learning classification. 
From this we can learn how host misidentification influences redshift assignment, photometric SN classification, 
and corrections for SN-host correlations and how these ultimately translate into biases in the derived cosmology.  
Additional study is required to determine what statistics are needed in order to replicate the conditions of the 
DES search for the purposes of simulation and training the methodology. 

\acknowledgments

RRG would like to thank Lindsey Bleem and Adrian Pope for helpful discussions regarding galaxy properties and catalogs. 

The submitted manuscript has been created by UChicago Argonne, LLC, Operator of Argonne National Laboratory (``Argonne"). 
Argonne, a U.S. Department of Energy Office of Science laboratory, is operated under Contract No. DE-AC02-06CH11357. The U.S. 
Government retains for itself, and others acting on its behalf, a paid-up nonexclusive, irrevocable worldwide license in said article to reproduce, 
prepare derivative works, distribute copies to the public, and perform publicly and display publicly, by or on behalf of the Government.
This research made use of Astropy, a community-developed core Python package for Astronomy \citep{astropy}. 
We acknowledge support from the MareNostrum supercomputer (BSC-CNS, www.bsc.es), 
Port d'Informac\'io Cient\'ifica (PIC, www.pic.es) and CosmoHUB (cosmohub.pic.es), 
where the MICE simulations were run, stored, and distributed, respectively.
M. Sullivan acknowledges support from EU/FP7-ERC grant no [615929].
Part of this research was conducted by the Australian Research Council Centre of Excellence for All-sky Astrophysics 
(CAASTRO), through project number CE110001020.

Funding for the DES Projects has been provided by the U.S. Department of Energy, the U.S. National Science Foundation, 
the Ministry of Science and Education of Spain, the Science and Technology Facilities Council of the United Kingdom, the 
Higher Education Funding Council for England, the National Center for Supercomputing Applications at the University of Illinois 
at Urbana-Champaign, the Kavli Institute of Cosmological Physics at the University of Chicago, the Center for Cosmology and 
Astro-Particle Physics at the Ohio State University, the Mitchell Institute for Fundamental Physics and Astronomy at Texas 
A\&M University, Financiadora de Estudos e Projetos, Funda{\c c}{\~a}o Carlos Chagas Filho de Amparo {\`a} Pesquisa do 
Estado do Rio de Janeiro, Conselho Nacional de Desenvolvimento Cient{\'i}fico e Tecnol{\'o}gico and the Minist{\'e}rio da Ci{\^e}ncia, 
Tecnologia e Inova{\c c}{\~a}o, the Deutsche Forschungsgemeinschaft and the Collaborating Institutions in the Dark Energy Survey.

The Collaborating Institutions are Argonne National Laboratory, the University of California at Santa Cruz, the University of Cambridge, 
Centro de Investigaciones Energ{\'e}ticas, Medioambientales y Tecnol{\'o}gicas-Madrid, the University of Chicago, University College 
London, the DES-Brazil Consortium, the University of Edinburgh, the Eidgen{\"o}ssische Technische Hochschule (ETH) Z{\"u}rich, 
Fermi National Accelerator Laboratory, the University of Illinois at Urbana-Champaign, the Institut de Ci{\`e}ncies de l'Espai (IEEC/CSIC), 
the Institut de F{\'i}sica d'Altes Energies, Lawrence Berkeley National Laboratory, the Ludwig-Maximilians Universit{\"a}t M{\"u}nchen 
and the associated Excellence Cluster Universe, the University of Michigan, the National Optical Astronomy Observatory, the 
University of Nottingham, The Ohio State University, the University of Pennsylvania, the University of Portsmouth, SLAC National 
Accelerator Laboratory, Stanford University, the University of Sussex, Texas A\&M University, and the OzDES Membership Consortium.

The DES data management system is supported by the National Science Foundation under Grant Number AST-1138766. 
The DES participants from Spanish institutions are partially supported by MINECO under grants AYA2012-39559, ESP2013-48274, 
FPA2013-47986, and Centro de Excelencia Severo Ochoa SEV-2012-0234.
Research leading to these results has received funding from the European Research Council under the European Union's Seventh Framework 
Programme (FP7/2007-2013) including ERC grant agreements 240672, 291329, and 306478.

\clearpage
\appendix
\section{GALAXY CLUSTERING: COMPARISON BETWEEN MICECATv2.0 AND ACS-GC COSMOS}
\label{Sec:Appendix}

In this appendix, we go into more detail about the differences between MICECATv2.0 and the ACS-GC 
COSMOS catalog we use in this work.  In an effort to better understand the reason why the host matching accuracy 
is lower for SNe simulated on MICE galaxies compared to those simulated on ACS-GC we examine the clustering 
properties of the two catalogs, particularly at the small scales we are concerned with in this work 
(i.e., $< 30$ arcsec). This comparison is done using only the positions of the galaxies (after a magnitude cut) 
and does not rely on their shapes or orientations.

First, we begin by attempting to make the two catalogs as similar as possible. We remove compact objects 
(defined in Section~\ref{Sec:ACSCat}) from the ACS-GC catalog, leaving only galaxies.  Then we impose a 
magnitude limit on both catalogs, requiring $i<24$ mag for MICE and MAG\_BEST (F814W) $<24$ mag for 
ACS-GC, where we expect both catalogs to be complete. 
Since the ACS F814W is a broad $i$ filter, not identical to DES $i$ band, this will result in minor differences. 
We then sample 10,000 random galaxies each from these magnitude-limited MICE and ACS-GC catalogs. 
For each of these randomly-selected galaxies we compute several quantities: the projected angular distance to the 
nearest neighbor and the number of other galaxies within radii of various sizes 
(30\arcsec, 20\arcsec, 10\arcsec, 5\arcsec, 2\arcsec, and 1\arcsec).

\begin{figure}[h]
\epsscale{0.6}
\plotone{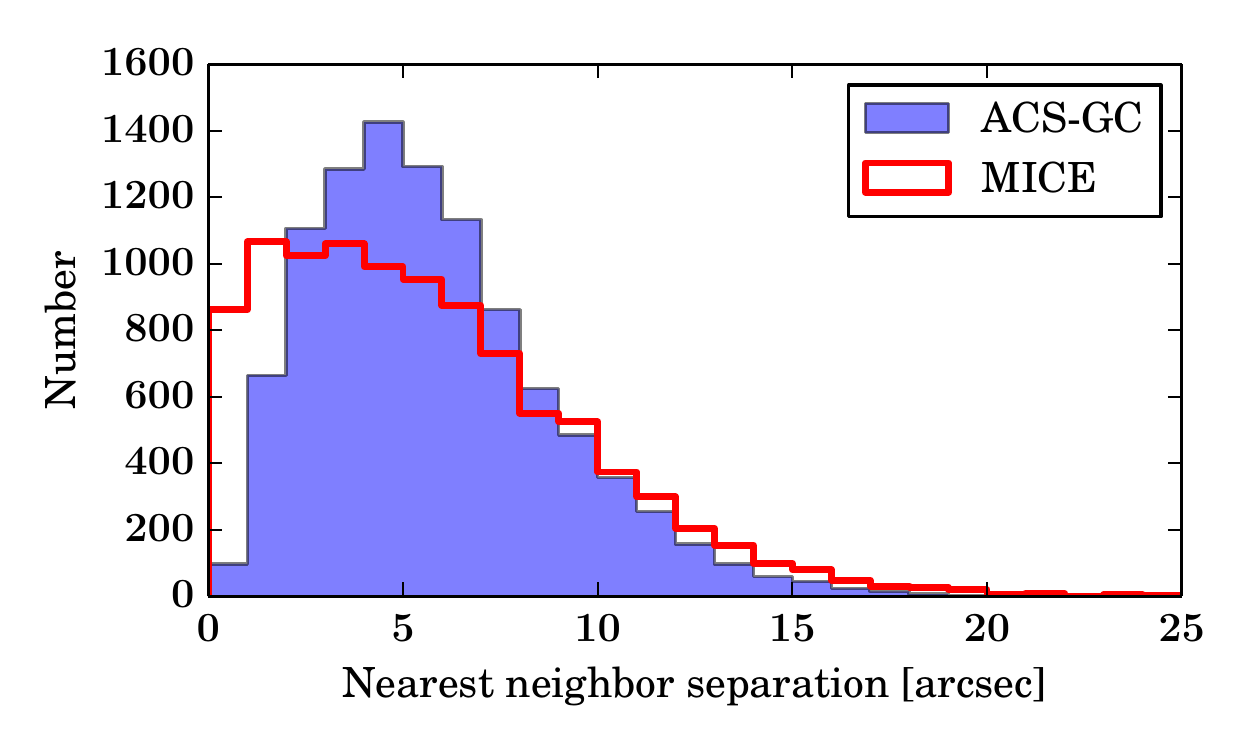}
\caption
{Distance to the nearest neighboring galaxy for a random sample of MICECATv2.0 galaxies and ASC-GC COSMOS 
galaxies. On scales smaller than $\approx 2$ \arcsec\ MICE exhibits a higher degree of clustering 
compared with ACS-GC data.}
\label{fig:NNsep}
\end{figure}

In Figure~\ref{fig:NNsep} we plot the distribution of nearest neighbor separations.  While the mean values of the 
distributions are quite similar (5.76\arcsec\ for MICE and 5.84\arcsec\ for ACS-GC), we see that the distributions 
themselves are quite different.  Particularly telling is the discrepancy below 2\arcsec\ in which we see 
that it is fairly common for MICE galaxies to have other galaxies very nearby ($\sim 20\%$ of MICE galaxies have 
neighbors within 2\arcsec), whereas such an occurrence in ACS-GC is rare. 
While the ACS PSF FWHM is very small (0.09\arcsec), it is possible that the deblending of galaxies within 2\arcsec\ 
is sometimes problematic in the \hst\ data.

\begin{figure}[tb]
\epsscale{1.1}
\plotone{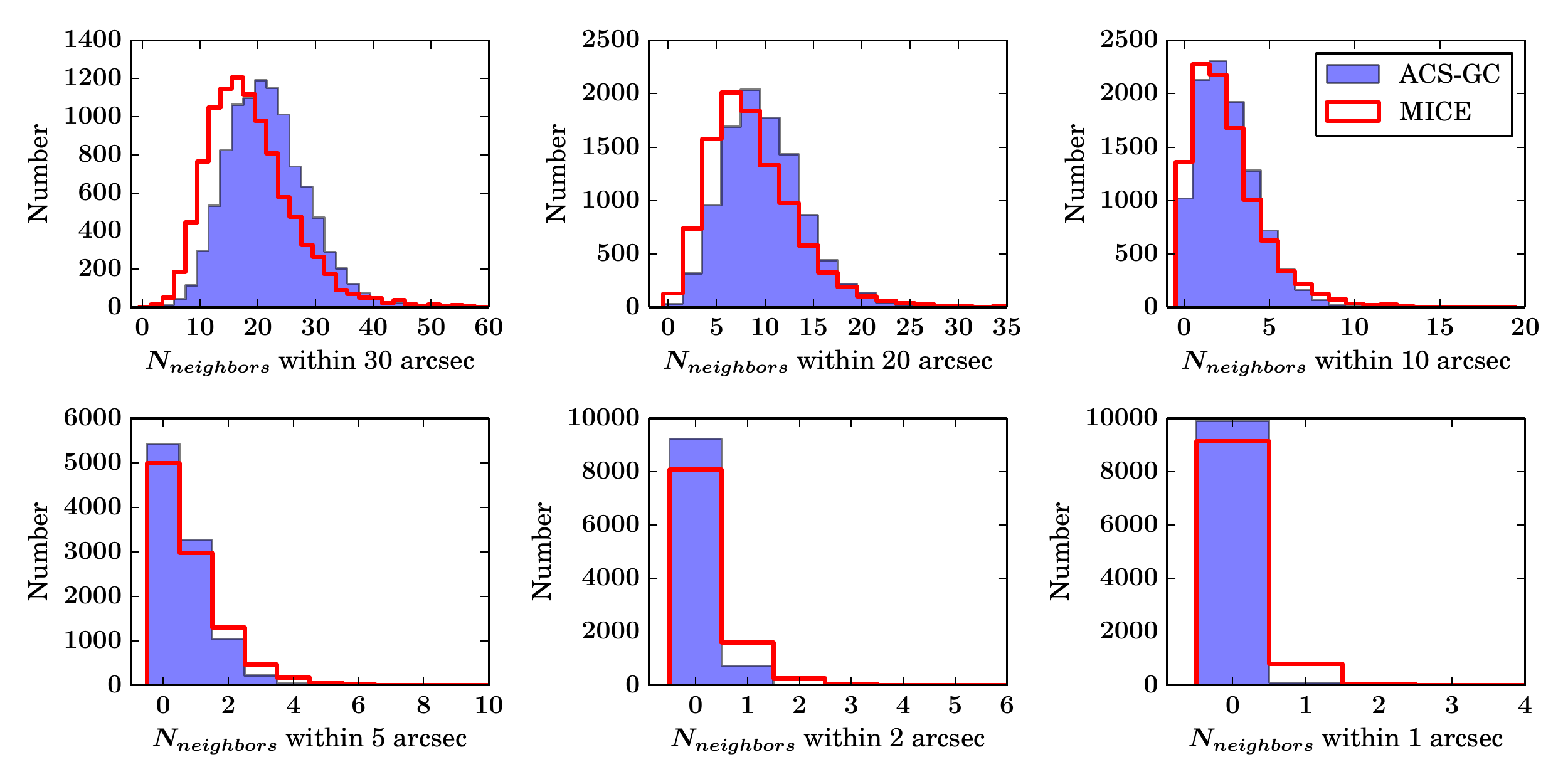}
\caption
{Distributions of the number of neighboring galaxies within various distances from random galaxies selected from 
MICECATv2.0 and ACS-GC COSMOS.}
\label{fig:Nwithin}
\end{figure}

In Figure~\ref{fig:Nwithin} we plot the distribution of the number of neighboring galaxies within 6 different radii. In the top 
panels (showing radii of 30\arcsec, 20\arcsec, and 10\arcsec), the ACS-GC distributions lie to the right of the 
MICE distributions, which indicates that when averaging over regions of this size, the ACS-GC catalog has a slightly 
higher mean galaxy density. However, when we examine regions of smaller area (such as in the lower panels showing 
radii of 5\arcsec, 2\arcsec, and 1\arcsec), we see 
the opposite effect: MICECAT has a higher mean galaxy density.  For example, the last panel in the lower right shows 
that a random galaxy in MICECAT has nearly a 10\% probability of having another galaxy within 1\arcsec, while for 
the ACS-GC catalog this probability is only 1\%.  
MICECAT was calibrated to reproduce the galaxy clustering observations at low redshift. In order to fit the clustering at 
small separations (the one-halo term), the galaxy distribution profile inside halos was made more concentrated than a 
standard NFW profile \citep{nfw97}.  The need for this extra concentration was extrapolated at higher redshift given the 
lack of calibrating data. Also, the galaxy mock generating code contains also a minimum radius for satellites inside their 
halos below which satellites are considered to have merged with the central halo. The extrapolation of the extra concentration 
at higher redshift and/or an underestimation of the minimum ``merging radius" used may contribute to the higher number 
of galaxy pairs seen in the simulation mock catalog compared to the ACS-GC data.

These differences in clustering properties between the MICECATv2.0 and ACS-GC COSMOS catalogs have  implications for 
our study of host galaxy matching since the probability of a SN being correctly matched to its host galaxy is highly dependent 
on the very local galaxy density. We have shown here that for MICECAT, the clustering on scales smaller than 5\arcsec\ is 
enhanced relative to ACS-GC. Further investigation of the subset of MICECAT galaxies with a neighbor within 2\arcsec\ shows 
that in two-thirds of these cases, the neighboring galaxy has a redshift within 0.0001 of the random galaxy's redshift; this 
indicates that they belong to the same halo and thus are true neighbors and not merely projected coincidences.  
In half of the cases where the neighbor lies within 2\arcsec, 
the galaxy and its neighbor overlap each other at the 1 half-light-radius level. This implies that roughly 
10\% of all MICECAT galaxies overlap with other galaxies. Since these are simulated galaxies, all of them appear in the mock 
catalog, whereas in a real catalog some of these would not be detected due to occlusion of galaxies along the line of sight or 
an inability to deblend overlapping galaxies.

This enhanced clustering in concert with the overlap issue in MICECAT would account for the overall lower matching
accuracy using MICECAT (90\%) compared to ACS-GC (92\%), since a higher local galaxy density increases the potential 
for confusion and mismatch. Most science being tested with mock catalogs of this kind (such as weak lensing or large 
scale structure studies) do not care about scales this small. 
Further studies are needed to determine if the clustering we see in MICECAT and ACS-GC on small scales 
is real or due to some deficit of simulations or deblending issue with actual data and source detection.

\clearpage
\bibliographystyle{/Users/ravigupta/Dropbox/aastex52/LFpaper/apjhack}
\bibliography{/Users/ravigupta/Dropbox/aastex52/LFpaper/myrefs}

\end{document}